\documentstyle[12pt]{article}

\def\be{\begin{equation}}
\def\ee{\end{equation}}
\def\bea{\begin{eqnarray}}
\def\eea{\end{eqnarray}}

\def\br{\begin{thebibliography}{abc}}
\def\er{\end{thebibliography}}

\def\inbar{\,\vrule height1.5ex width.4pt depth0pt}
\def\IB{\relax{\rm I\kern-.18em B}}
\def\IC{\relax\hbox{$\inbar\kern-.3em{\rm C}$}}
\def\ID{\relax{\rm I\kern-.18em D}}
\def\IE{\relax{\rm I\kern-.18em E}}
\def\IF{\relax{\rm I\kern-.18em F}}
\def\IG{\relax\hbox{$\inbar\kern-.3em{\rm G}$}}
\def\IH{\relax{\rm I\kern-.18em H}}
\def\II{\relax{\rm I\kern-.18em I}}
\def\IK{\relax{\rm I\kern-.18em K}}
\def\IL{\relax{\rm I\kern-.18em L}}
\def\IM{\relax{\rm I\kern-.18em M}}
\def\IN{\relax{\rm I\kern-.18em N}}
\def\IO{\relax\hbox{$\inbar\kern-.3em{\rm O}$}}
\def\IP{\relax{\rm I\kern-.18em P}}
\def\IQ{\relax\hbox{$\inbar\kern-.3em{\rm Q}$}}
\def\IR{\relax{\rm I\kern-.18em R}}
\font\cmss=cmss10 \font\cmsss=cmss10 at 7pt
\def\IZ{\relax\ifmmode\mathchoice
{\hbox{\cmss Z\kern-.4em Z}}{\hbox{\cmss Z\kern-.4em Z}}
{\lower.9pt\hbox{\cmsss Z\kern-.4em Z}}
{\lower1.2pt\hbox{\cmsss Z\kern-.4em Z}}\else{\cmss Z\kern-.4em Z}\fi}
\def\IGa{\relax\hbox{${\rm I}\kern-.18em\Gamma$}}
\def\IPi{\relax\hbox{${\rm I}\kern-.18em\Pi$}}
\def\ITh{\relax\hbox{$\inbar\kern-.3em\Theta$}}
\def\IOm{\relax\hbox{$\inbar\kern-3.00pt\Omega$}}

\def\rt{\rightarrow}

\def\bar#1{\overline{#1}}

\def\Hat#1{\rlap{\kern.10em$\widehat{\phantom G}$}#1}
\def\HAt#1{\rlap{\kern.05em$\widehat{\phantom G}$}#1}

\def\czp#1{\rlap{\kern.1em$\widehat{\phantom{G\vrule height.8em}}$}#1{}}
\def\Czp#1{\rlap{\kern.05em$\widehat{\phantom{G\vrule height.8em}}$}#1{}}

\def\semidirect{\mathbin
               {\hbox
               {\hskip 3pt \vrule height 5.2pt depth -.2pt width .55pt 
                \hskip-1.5pt$\times$}}}
\newcommand{\sect}[1]{\setcounter{equation}{0}\section{#1}}

\def\sxn#1{\bigskip\medskip \sect{#1} \smallskip
                                                 }

\begin{document}
\input epsf.tex
\thispagestyle{empty}
\setcounter{page}{0}

\begin{flushright}
SU-4240-702\\
\end{flushright}

\vspace{1cm}

\begin{center}
{\LARGE The Spin-Statistics Connection in Quantum Gravity  \\}
\vspace{5mm}
\vspace{2cm}
{A.P. Balachandran$^1$, E. Batista$^2$,  I.P. Costa e Silva$^3$ and 
P. Teotonio-Sobrinho$^3$}\\

\vspace{.50cm}
$^1${\it Department of Physics,Syracuse University \\
Syracuse, NY 13244-1130, USA}\\
\vspace{.50cm}
$^2${\it Universidade Federal de Santa Catarina, Centro de F\'isica e
  Matem\'atica,\\
Dep. MTM, CEP 88.010-970, Florian\'opolis, SC, Brazil}\\
\vspace{.50cm}
$^3${\it Universidade de Sao Paulo, Instituto de Fisica-DFMA \\
Caixa Postal 66318, 05315-970, Sao Paulo, SP, Brazil}\\
\vspace{2mm}
\vspace{.2cm}

\end{center}

\begin{abstract}
It is well-known that in spite of sharing some properties with
conventional particles, topological geons in general violate the
spin-statistics theorem. On the other hand, it is generally believed that
in quantum gravity theories allowing for topology change, using pair
creation and annihilation of geons, one should be able to
recover this theorem. In this paper, we take an alternative route, and
use an algebraic formalism developed in previous work. We give a
description of topological geons where an algebra of
``observables'' is identified and quantized. Different irreducible
representations of this algebra correspond to different kinds of
geons, and are labeled by a non-abelian ``charge'' and ``magnetic
flux''. We then find that the usual
spin-statistics theorem is indeed violated, but a new spin-statistics
relation arises, when we assume that the fluxes are
superselected. This assumption can be proved if all observables are
local, as is generally the case in physical theories. Finally, we
 also discuss how our approach fits into conventional formulations of
 quantum gravity.
\end{abstract}

\newpage

\sxn{Introduction}\label{S1}

The spin-statistics theorem is one of the most fundamental relations
in the theories describing the particles of nature. As far as experimental
tests are concerned, no elementary particles were ever found which violate it.
It was therefore a surprise when it was discovered \cite{erice,spin-stat}
in the middle 80's that topological geons did violate this
relation in general, at least in the case when the spatial topology is
not allowed to change. Topological geons are soliton-like 
excitations of the spatial manifold $\Sigma$ \cite{erice,FS}. They can
be thought of as lumps of nontrivial topology. 
For example, in \mbox{$(2+1)d$}, the
topology of an orientable, closed surface $\Sigma $ is determined by the
number of connected components of $\Sigma $ and by the number of
handles on each connected component. Each handle corresponds to
a topological geon, i.e., a localized lump of nontrivial topology. 
It is well known that these
solitons have particle-like properties such as spin and statistics.
However, as we observed before, unlike ordinary particles they can
violate the  spin-statistics 
relation. It has been suggested
\cite{erice,spin-stat,DS,where} that the standard spin-statistics
relation can be recovered if one considers processes where geons are
(possibly pairwise) created and annihilated, but this necessarily
implies a change of the topology of $\Sigma$. In other words, one may
have to consider topology change in order to have a spin-statistics
theorem for geons \cite{DS,where}. 

To appreciate the importance of having (or not having) a spin-statistics
connection for geons, one must recall that in ordinary
quantum field theories in Minkowski space, the particles which arise
when we second quantize, for instance, have this connection
naturally. Now, in a hypothetic quantum theory of topology, one could think of
geons as a kind of ``particle'', representing excitations of the
topology itself. It seems therefore natural to ask about whether
they share this connection with ``true''particles. As we have mentioned,
they do not, but still we find that in the formalism we develop here
a different, weaker version of the spin-statistics connection arises.

In the absence of a full-fledged quantum gravity theory, it has become
a current practice to consider simple models which retain some of its aspects
while being more tractable in the formal aspects. Accordingly, our intention in
this work is to use a very simple model, a gauge theory with a finite
gauge group in $(2 + 1)d$ space-time dimensions, to understand the
spin-statistics theorem in quantum gravity. This model has the
advantage of ``isolating'' the topological degrees of freedom, which
are in a certain sense canonically quantized independently from
degrees of freedom coming from metric and other fields. The same model
has been considered in a companion paper \cite{us}, and there we show
that we may consider topology change as a quantum phenomenon depending
on the scale of observations. Therefore this model features spatial
topology change in some sense.  
Actually, in spite of the fact that topology change has been
inspired by quantum gravity, it has been demonstrated  in \cite{TC}
that it can happen in ordinary quantum mechanics. In this approach,
metric is not dynamical, 
but degrees of freedom related to topology are quantized. The notion of a 
space with a well defined topology appears only as a classical limit.
(See also \cite{marolf} for related ideas). 

Let us consider a manifold $M$ and some generic field theory 
(possibly with  gauge and Higgs fields) 
interacting with gravity. It is reasonable to expect that if we 
could quantize such a complex theory, its observables would 
give us information on the geometry and 
topology of $M$. The main point is that one does not need to consider the
full theory to get some topological information. It is possible
that, in a certain low energy (large distance) limit,  there would be
a certain set of  observables encoding the topological data. 
We know examples where this is precisely the case. In general, the 
low energy (large distance) limit of a field theory is not able to probe
details of the short distance physics, but it  can isolate degrees of
freedom related to
topology. We may give as an example the low energy limit of $N=2$
Super Yang-Mills,
known as the Seiberg-Witten theory \cite{SW}. We also have examples of
more drastic reduction where a field theory in the vacuum state 
becomes purely topological \cite{vacuum}. Inspired by these facts 
we will identify the degrees of freedom, 
or the algebra ${\cal A}^{(n)}$ of ``observables'',  
capable of describing $n$ topological geons in $(2+1)d$. Actually, we
will argue later in this paper that not all the operators in this algebra are
observables in the strict sense. Rather, this algebra is a
sort of {\it field algebra} (see, for
instance, \cite{Haag} and references therein for more information on
field algebras).  
We say that ${\cal A}^{(1)}$ 
describes a single
geon  in the same way that the algebra of angular momentum describes a
single spinning particle. In this framework what we mean by quantizing
the system is nothing but finding  irreducible representations
of ${\cal A}^{(1)}$. As in the case of the algebra of angular momentum,
different irreducible representations have to be thought of as
different {\it particles}.  
For the moment, we will not be concerned with dynamical aspects. 
We would like to
concentrate on the quantization itself and leave the dynamics to be
fixed by the particular model one wants to consider.  

An intuitive way of understanding the algebra ${\cal A}^{(1)}$ 
for a topological geon comes from considering a gauge
theory with
gauge group $G$ in two space dimensions spontaneously broken to a
discrete group $H$. For simplicity we will assume that $H$ is
finite. As an immediate consequence it follows that 
the gauge connection (at far distances) is locally flat. In other
words, homotopic (based) loops $\gamma $ and $\gamma '$ produce the same 
parallel transport (holonomy). The set of
independent holonomies are therefore parametrized by elements 
$[\gamma ]$ in the fundamental group $\pi _1(\Sigma )$.  
It is quite clear that such quantities are enough
to detect the presence of a handle. The phase space we are 
interested in contains only topological degrees of freedom. Therefore 
such holonomies can be thought of as playing the role of position variables. 
We also have to take into account the 
diffeomorphisms (diffeos) that are able to change $[\gamma ]$. They will be
somewhat the analogues of translations. 
It is clear that the connected component of the group of
diffeomorphisms preserving the basepoint and a frame thereat, the
so-called small diffeos, cannot
change the homotopy class of $\gamma $. 
To change the homotopy class of
a curve $\gamma $ one needs to act with the so-called large
diffeomorphisms. 
Therefore the analogues of
translations have to be parametrized by the large diffeos modulo the
small diffeos. This is exactly the mapping class group
$M_\Sigma $. Also, since we must fix a base point $P$ to define
$\pi_1(\Sigma)$, we must take into account the fact that the discrete
group $H$ can change the holonomies by a conjugation. These three sets of
quantities will
comprise our algebra ${\cal A}^{(1)}$. Contrary to what happens in field
theory or even in quantum mechanics, we find that ${\cal A}^{(1)}$ is finite 
dimensional. This will be important to avoid technical problems of
various kinds. The algebra ${\cal A}^{(1)}$ contains the analogue of  
positions and translations and can be thought as
a discrete Weyl algebra. There seems to be no great obstacle to generalize our
results also to the case where $H$ is a Lie group \cite{gen}.

Our algebraic description of geons is analogous to what has been developed
for  $2d$ non-abelian vortices by the Amsterdam group
\cite{prop}. These ideas have been further developed by some of us and
coworkers and applied to rings in $(3+1)d$. Their results will not be
discussed here since a complete account will be reported in \cite{syr}.

The algebra encountered by \cite{prop} was a special type of Hopf
algebra, namely the Drin'feld double of
a discrete group \cite{kassel}. In our case, however, 
the algebra ${\cal A}^{(1)}$ is not Hopf, but it has  a Drin'feld double
as a subalgebra. For a pair of geons we find that the corresponding
algebra ${\cal A}^{(2)}$ is 
closely related to the tensor product 
${\cal A}^{(1)}\otimes {\cal A}^{(1)}$ of
single geon algebras. This fact allows us to determine
the appropriate algebra ${\cal A}^{(n)}$ for an arbitrary number $n$ of
geons. Among the elements of ${\cal A}^{(2)}$ we find the elements
corresponding to the operations of exchanging the positions of two geons
and  rotating one of them  by $2\pi $. These are the two operations we
need in order to answer whether there is a spin-statistics relation. The usual
theorem states that the exchange of two identical components (statistics) 
is equivalent to the rotation by $2\pi $ of one of its components. It turns
out that this is no longer true. However, spin and statistics
are not independent but fulfill a well-defined relation. 

We would
like to point out some differences with respect to the approaches of
\cite{spin-stat} for the spin-statistics connection. To show their results,
the authors of \cite{spin-stat} have used anti-particles together
with rules for pair creation and annihilation. 
In our approach the spin-statistics relation follows entirely from the
the properties of the field algebras. It is true that we can
also have creation and annihilation of geons, but these processes are
not directly linked to the spin-statistics relation. For other approaches
to the spin-statistics theorem see \cite{BR,Anandan,DSu}. 

One advantage of the algebraic approach is that we can do this
analysis without going into the details of the ``complete'' underlying
field theory. We can determine the  spectrum ${\hat {\cal A}^{(1)}}$ of the 
geons, i.e., the set of possible irreducible representations of ${\cal
  A}^{(1)}$, but a particular field theory may restrict the available
possibilities in ${\hat {\cal A}^{(1)}}$. The determination of these
possibilities 
requires the study of  particular examples of the underlying 
field theories. That may be a very difficult task. In this paper our intention
is to use the simplified algebraic ``field'' theory and see what it can teach
us. It is remarkable that such a simple framework can reveal important
features of quantum geons such as a constraint involving spin and statistics. 
Rules for quantum topology change are discussed in a companion paper
\cite{us}.  

There is a systematic way to incorporate our algebraic methods in
conventional approaches to quantum gravity. When that is done, we end
up selecting a particular class of vector bundles, the sections of
which are state vectors of quantum gravity (they specify domains of
operators like the Hamiltonian). We shall discuss these issues in
detail elsewhere, limiting ourselves to a concise discussion in
Section \ref{S7} in this paper. The present paper therefore can be
interpreted as 
sponsoring the use of these bundles in quantizing gravity. We think
that there are powerful reasons supporting this point of view. Indeed
our work here shows that these bundles nicely incorporate information
on classical spatial topology and imply a (generalized)
spin-statistics theorem, whereas if this selection of bundles is
abandoned, there are many possible choices of bundles in the presence
of geons, and most do not imply any sort of spin-statistics
connection.

This paper is organized as follows. The field algebras
${\cal A}^{(n)}$ are described in Section \ref{S3}. 
In particular, the representations of 
${\cal A}^{(1)}$ will play an important role when we discuss the
spin-statistics connection. Quantization of
the system is given in Section \ref{S4}. In this section we are able
to classify the irreducible representations for a class of algebras
${\tilde {\cal A}}$ that includes our algebra of interest as a
particular example. It is important to note that these Sections are
shortened copies of Sections in the companion paper \cite{us}, which
we reproduce here for the benefit of the reader, rendering this paper
basically self-contained. The original part is in the following Sections. 
The existence of a novel spin-statistics connection for $2d$ 
orientable geons is established in Section \ref{S6}, under certain
asumptions which become clear in Section \ref{Newsection}, with the
introduction of the property of clustering for a system of $N$ geons,
and superselection of the global fluxes of geons. Section \ref{S7}
explores how one can use the representations of the algebra of
observables for geons to obtain geon states in quantum gravity. The paper ends
with some general remarks and an outlook on future work.

\sxn{The Algebras for $(2+1)d$ Topological Geons }\label{S3}

Throughout this work our setting is a space-time of the form $\Sigma
\times \IR$. Here, the spatial manifold $\Sigma$ is two-dimensional,
and will be typically assumed to be a plane with one or several
handles. Topological geons in this $(2 + 1)d$ context are simply (for
orientable space-times) these handles on the spatial manifold (for a
more detailed account and a more general definition of geons see, for
instance, \cite{erice,us,Aneziris}). Our aim in this section is to define
some ``observables'' which describe the topological character of a geon.
As we will see later, the algebras we obtain contain some operators
which are not really observables, since they represent non-local
operators which are in a certain sense ``gauged away'' whenever we
perform physical measurements for geons. Thus we will refer to the kind
of algebra we will encounter as a {\em field algebra} \cite{Haag}.

The presentation of the field algebra of geons given here will not be
detailed. The reader is refered to \cite{us} for a more comprehensive
discussion.  

We will follow an approach  inspired by the work of the 
Amsterdam group, which is reported in ref. \cite{prop}. In this work, 
the group investigates the properties of vortex solutions of a 
$(2+1)d$ gauge field theory in Minkowski spacetime
where the gauge symmetry of a Lie group $G$ is spontaneously 
broken to a finite group $H$ by a non-vanishing expectation value of a
Higgs field $\Phi$. See \cite{prop} for details. The Lagrangian is given by 
\be
{\cal L} \, =\, \frac{1}{4} F^{a}_{\mu\nu} F^{\mu\nu}_{a} \, +\, 
Tr[(D_{\mu} \Phi)^{*} \cdot (D^{\mu} \Phi )] \, -\, V(\Phi ) \; ,\label{higgs} 
\ee
where $\mu ,\nu =0,1,2$, and $a$ is a Lie algebra index. For
simplicity, we assume that $G$ is connected and simply connected.
The fields $F^{a}_{\mu\nu}$ are the 
components of the field strength of the Yang-Mills potential 
$A_{\mu}^{a}$ and $D_{\mu}$ denotes the covariant derivative
determined by this potential. The Higgs field $\Phi$ is in the adjoint
representation and can be expanded in
terms of generators $T^a$ of the Lie algebra of $G$, and $V(\Phi )$ is
a $G$-invariant potential.
In this paper we shall be concerned with the low energy, or
equivalently, the long range behavior of this theory, in the
temporal gauge $A_{0}^{a} = 0$. This is obtained
by  minimizing the three
terms in the energy density separately. Minimizing the term
corresponding to the energy density of the Yang-Mills field, we obtain
the condition $F^{a}_{\mu\nu} = 0$, from which we conclude that we
are dealing only with flat
connections. The minimum of the potential restricts the values of the
Higgs field to the vacuum manifold, which is invariant
by $H$. Finally, the condition $D\Phi =0$, required for minimizing the
energy density from the second term, tells us that the holonomies
\be
\tau (\gamma )= 
P\exp \{ \int_{\gamma} A^{a}_{i} T_{a} ds^i \} \; ; \; i \in \{1,2\} \; ,
\label{flat}
\ee
take values in the finite group $H$. 

Here and in what follows we will
fix a base point $P$ for loops, so that all loops will begin and end
at $P$.    

This gauge theory may have topologically non-trivial, static solutions
such as vortices. It is very well known that the core radii of these
vortices are inversely proportional to the mass of the Higgs boson,
and therefore they may be viewed as point-like in the low-energy
regime of the theory. Hence, according to a standard argument, to
describe the $N$-vortex solutions we may consider solutions for
the vortex equations 
\bea
\label{static}
F^{a}_{ij} &=& 0 ; \nonumber \\
D_i\Phi &=& 0 ; \nonumber \\
V(\Phi) &=& 0,
\eea
on a spacetime of the form $\Sigma \times \IR$, where $\Sigma$ is the
plane with $N$ punctures, playing the role of the vortices. 

One way to explain our approach is based on a field theory like
(\ref{higgs}). Addition of gravitational terms to (\ref{higgs}) would not
affect our arguments. The
difference in our approach is that we shall work in the zero vortex
number sector of this theory, but on a plane with geons. Hence, instead of
puncture, the non-trivial topology is characterized by
handles. Now, take
a solution $(A,\Phi)$ for the vortex equations (\ref{static}). By fixing a
point $P \in \Sigma$, the holonomy of $A$ around any path $\gamma$
based at $P$ depends only on its homotopy class, since $A$ is flat.
It takes values in a subgroup $H$ of $G$, which preserves the vacuum
manifold, in view of the equations for $\Phi$ \cite{prop}. Therefore,
any solution of the vortex equations determines a
homomorphism $\tau$,
\be
\label{homo}
\tau :\pi _1(\Sigma ) \rt H,
\ee
of the fundamental group $\pi_1 (\Sigma )$ to the group
$H$. Conversely, given such a homomorphism $\tau$ we can define a
solution for eqs.(\ref{static}) in the following way. Take the
universal covering space $\tilde{\Sigma}$ of $\Sigma$. It is the total
space of a principal bundle over $\Sigma$ with structure group $\pi_1
(\Sigma)$. Via the homomorphism $\tau$ we can construct an associated
principal $H$-bundle over $\Sigma$, which is a subbundle of the
original $G$-bundle. Since $H$ is finite, this bundle
has a unique flat connection $A^{a}_{i}$, which can be viewed as a reducible
connection on the $G$-bundle. We now find a $\Phi$. After
fixing some $\Phi _0$ in the vacuum manifold, we define $\Phi(P) =
\Phi _0$. Then, since
$\Phi$ must be covariantly constant, its value can be obtained for
each $x \in \Sigma$ by parallel transporting $\Phi _0$ along some path
from $P$ to $x$ in $\Sigma$: 
\be
\label{transport}
\Phi(x)= 
P\exp \{ \int_{P}^{x} A^{a}_{i} T_{a} ds^i \} \Phi _0. 
\ee
The pair ($A^{a}_{i}$, $\Phi$) thus constructed is obviously a
solution of the vortex equations. Therefore the space of solutions for
the vortex eqs. (\ref{static})
is essentially parametrized by homomorphisms $\tau : \pi_1(\Sigma)
\rightarrow H$. Each such homomorphism is then a vortex configuration
when we have punctures. In our context, we will call one such
homomorphism a geon configuration. In general, it gives non-zero
``magnetic fluxes'' around non-trivial
elements of $\pi_1(\Sigma)$.

The finite group $H$ acts on the space of solutions. In terms of homomorphisms
we have that, under these $H$-transformations, a flux
$\sigma$ transforms as
\be
\sigma\, \mapsto \, h \sigma h^{-1} \; ,\label{gauge}
\ee
In other
words, we have an action of $H$ by conjugation of the fluxes. We shall
simply refer to this action as the $H$-transformations. The group
elements $h \in H$ will be regarded as operators when we quantize the
theory, also denoted
by $h$. The multiplication of two $H$-transformations is the same as the
group multiplication. Therefore the algebra of such 
operators turns out to be the group algebra ${\IC }(H)$. 

As for the physical interpretation of the
$H$-tranformations we note that the mathematical action depicted in
(\ref{gauge}) is entirely equivalent, from a physical standpoint, to
what occurs when one makes a flux $\sigma$ encircle a source
of flux $h$ at infinity. Since such an operation is non-local, one must
conclude that the $H$-transformations cannot be
considered local in the theory, i.e., cannot be implemented by local
operators.  

The total algebra in the case of vortices (punctures)is the
semi-direct product $D(H)={\IC}(H)\semidirect {\cal F}(H)$, where
${\cal F}(H)$ is the algebra of complex-valued functions on $H$ with
product given by pointwise multiplication: it
describes the ``position observables'' for a vortex. The reason
for this denomination will become clear when we discuss geons. The algebra
$D(H)$ is the so-called Drin'feld double \cite{prop}. It has the
structure of a quasi-triangular Hopf algebra. The Hopf structure \cite{kassel} 
means in particular the existence of a co-product, i.e, a map 
\[
\Delta \, :\, D(H)\,  \longrightarrow \,  D(H)\otimes D(H) \; ,
\]
which is a homomorphism of algebras. In \cite{prop} the fluxes are
seen as particles in $(2 + 1)d$ and are then first quantized: the (internal)
Hilbert space ${\cal H}$ is constructed, and the
elements of the algebra $D(H)$ act as operators on this Hilbert space.
${\cal H}$ decomposes into irreducible representations of $D(H)$,
corresponding to the different particle sectors of the quantum
theory. The existence of a co-product allows one to understand fusing processes between
particles. The quasi-triangularity implies the existence of the
$R$-matrix, $R \in D(H)\otimes D(H)$, responsible for all braiding 
processes between particles. Again, for further details see \cite{prop}. 

How is the topology of $\Sigma$ taken into account in this
approach? First of all, we have seen that the physically distinct
(for vortices and/or geons) configurations are in one-to-one
correspondence with conjugacy classes of homomorphisms of $\pi_1(\Sigma)$ into
$H$. Moreover, it is well known that for a finite group $H$ the
elements in the latter
space are in one-to-one correspondence with equivalence classes of
principal $H$-bundles over $\Sigma$. Therefore the only degree of
freedom in this theory is the topology of these bundles. Second, a
configuration for which the holonomy is
trivial around some puncture or handle is indistinguishable, from the
standpoint of the low-energy theory, to another in which that
particular puncture or handle
is absent. Therefore the low-energy theory somehow actually allows for
``topology fluctuations'' of $\Sigma$ as long as we stay within its
limits. Such a way of viewing topology change is explored in
\cite{us}. It is very much akin to the views pursued in
non-commutative geometry, where one uses an algebra to encode
space-time geometry and topology. In this approach the usual
``classical'' view of a background manifold is secondary, and the
topology is actually viewed as a consequence of the algebraic setting
one uses. 

In order to determine the field algebra for a topological 
geon, we will first try to find the analogues of the ``position 
observables'' for a geon. Now, $\Sigma$ is the plane with one or more
handles, and for simplicity we shall assume throughout that there
are no vortices, i.e., we work in the zero vortex number sector of the
low-energy limit of the theory given by the Lagrangian in
(\ref{higgs}). In this case, all non-trivial configurations will be
related solely to holonomies around and through the handles.    

\begin{figure}[t]
\begin{center}
\begin{picture}(0,0)
\put(90,-10){$P$}
\put(93,2){$\ast$}
\put(50,50){$\gamma_1$}
\put(100,50){$\gamma_2$}
\put(190,50){$\gamma_3$}
\end{picture}
\epsfbox{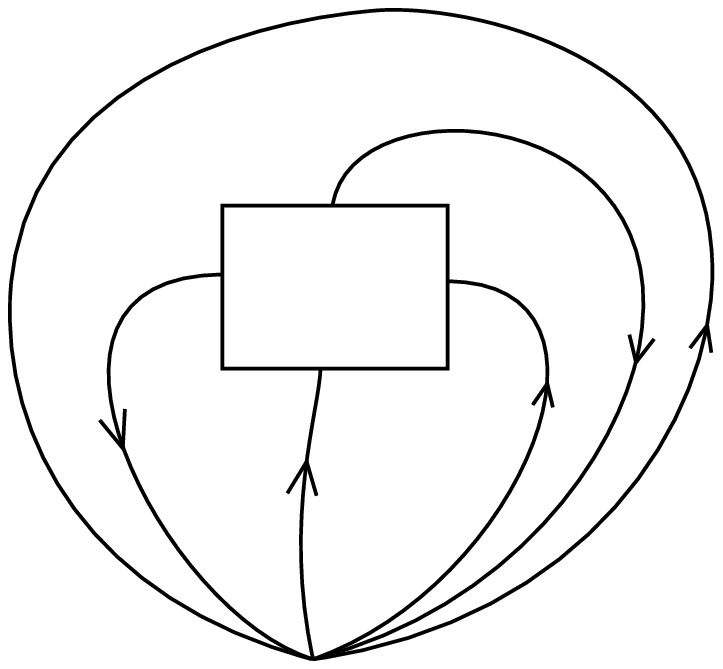}
\end{center}
{{\bf Fig. 2.1:} The figure shows the loops $\gamma_i$ ($1 \leq i \leq
  3$). The homotopy classes $[\gamma_1]$ and $[\gamma_2]$ generate the
  fundamental group. The class $[\gamma _3]$ is not independent of
  $[\gamma_1]$ and $[\gamma_2]$.}
\end{figure} 

Let us start by taking $\Sigma $ to be the plane with a handle. On all
figures, a geon will be thought of as a square hole on the plane,
with the opposite sides identified. One can show that 
$\pi _1(\Sigma )$ has two generators $[\gamma _1]$ and $[\gamma _2]$, shown by 
Fig. 2.1. It can be shown that 
\[
[\gamma _3]=[\gamma_1][\gamma _2][\gamma_1]^{-1}[\gamma _2]^{-1}
\]
Actually, $\pi _1(\Sigma )$ is freely generated by
$[\gamma _1]$ and $[\gamma _2]$. Let 
$g=W([\gamma _1],[\gamma _2])\in \pi _1(\Sigma ) $, be a word in 
$[\gamma _1],[\gamma _2]$ and their inverses.
Then $\tau $ maps $g$ to  
$W(a,b)\in H $ where 
$a=\tau (\gamma _1)$ and $b=\tau (\gamma _2)$.  Therefore the map 
$\tau :\pi _1(\Sigma )\rt H$ is completely 
characterized by the fluxes $\tau (\gamma _1)=a$ and 
$\tau (\gamma _2)=b$. Since there is no relation between $a$ and $b$, the set 
$T$ of all maps is labeled by $H\times H $. 

We now define more precisely what we mean by a geon configuration. Let $H$ be a finite group and $\Sigma$ the plane with one 
geon, i.e., a two dimensional  manifold given by 
\[
M\, =\, {\bf R}^2 \, \# \, {\bf T}^2 \; .
\]
Let $\gamma _1$ and $\gamma _2$ denote representative loops whose classes 
generate $\pi_1 (\Sigma )$. 
We define a  classical configuration $\tau _{(a,b)}\in T$ 
of a geon as the homomorphism defined by
\be
\tau _{(a,b)} (\gamma _1)=a,\mbox{ and } \tau _{(a,b)}(\gamma _2)=b.
\ee

It is important to bear in mind that $T \cong H\times H$ and therefore 
that it is a finite discrete set. 
For simplicity of notation, a geon configuration will be denoted 
simply by a pair $(a,b)$ of fluxes. Note that we are not explicitly
identifing those configurations which differ by an
$H$-transformation. This is because wave functions need only be
``covariant'' under the symmetries of the problem, and only its
modulus squared and other observable quantities, like Aharonov-Bohm
phases, must be invariant. In our approach, this will happen
naturally, just as in \cite{prop}.    

With $T\cong H\times H$ being the configuration space for a geon, the 
corresponding algebra of ``position observables'' is ${\cal
  F}(T)$, the algebra of complex-valued functions on $T$ with product
given by pointwise multiplication. Instead of working with the
abstract algebra, we specify a
representation. Let $V$ be the (finite-dimensional) complex vector
space generated by the vectors    
$\mid a,b\rangle , a,b \in H$. We will call the representation on
$V$, to be defined below, the defining representation.
The algebra ${\cal F}(T)$ 
is generated by projectors on $V$ denoted by $Q_{(a,b)}$. They
are defined by  
\be
\label{projectors}
Q_{(a,b)} \mid c,d \rangle \, =\, \delta_{a,c} \, \delta_{b,d} 
\mid c,d \rangle \; .
\ee
The operator $Q_{(a,b)}$ represents a ``delta 
function'' supported at $(a,b)$, i.e., it gives $1$ when evaluated on $(a,b)$,
and zero everywhere else. Indeed, from (\ref{projectors}) one finds that
\be
\label{algebraF}
Q_{(a,b)} Q_{(c,d)} \, =\,  \delta_{a,c} \delta_{b,d}  Q_{(c,d)} \; .
\ee

Besides the projectors $Q_{(a,b)}$, which play the role of
position operators in ordinary quantum mechanics, we have also some
operators capable of changing $(a,b)$. They are somewhat 
analogous to momentum operators. For example, like in the case of vortices,  
$H$-transformations act on the configurations. 
It turns out that for a geon there are additional operators besides 
$H$-transformations. They correspond to the 
action of the group  $Diff ^\infty(\Sigma ) $ 
of diffeomorphisms of $\Sigma $ that keeps infinity invariant. 

We will start by first examining the $H$-transformations. 

The group $H$ acts on $T$ simply by conjugating both fluxes in $(a,b)$. This 
will induce an operator $\hat \delta _g$ for each $g\in H$, acting on the defining 
representation $V$ by 
\be
\hat \delta _g \mid a,b \rangle \, =\, \mid gag^{-1} ,gbg^{-1} \rangle. 
\label{del}
\ee
From (\ref{del}) one sees that the multiplication of operators 
$\hat \delta _g$ is given by 
\be
\label{groupalgebra}
\hat \delta_{g} \hat \delta_{h} \, =\, \hat \delta_{gh} \; .
\ee
The corresponding algebra generated by $\delta _g$ is the group algebra 
${\IC}(H)$. The relation between ${\cal F}(H\times H)$ and ${\IC}(H)$ 
can be derived from (\ref{projectors}) and (\ref{del}). One sees immediately
that 
\be
\label{actionofH}
\hat \delta _g Q_{(a,b)}\hat\delta ^{-1} _g \, =
\, Q_{(gag^{-1},\, gbg^{-1} )} \; .
\ee
In other words, the algebra ${\IC}(H)$ acts on ${\cal F}(H\times H)$.

Besides $H$-transformations, fluxes $(a,b)$ can change under the action of 
the group $Diff ^{\infty}(\Sigma )$. 
It is clear that elements belonging to the subgroup 
$Diff ^{\infty}_0 (\Sigma )$, the component connected to identity,
 act trivially on $\pi _1(\Sigma )$ \footnote{For simplicity, we take
the basepoint $P$ to be at infinity.} and hence on 
$(a,b)$. Therefore what matters is the action of the so-called mapping
class group $M_\Sigma $ \cite{birman,balachandran}, defined as
\be
\label{mcg}
M_{\Sigma } \, =\, 
\frac{Diff^{\infty} (\Sigma )}{Diff^{\infty}_{0}(\Sigma )} \; .
\ee
\begin{figure}[t]
\centerline{\epsfbox{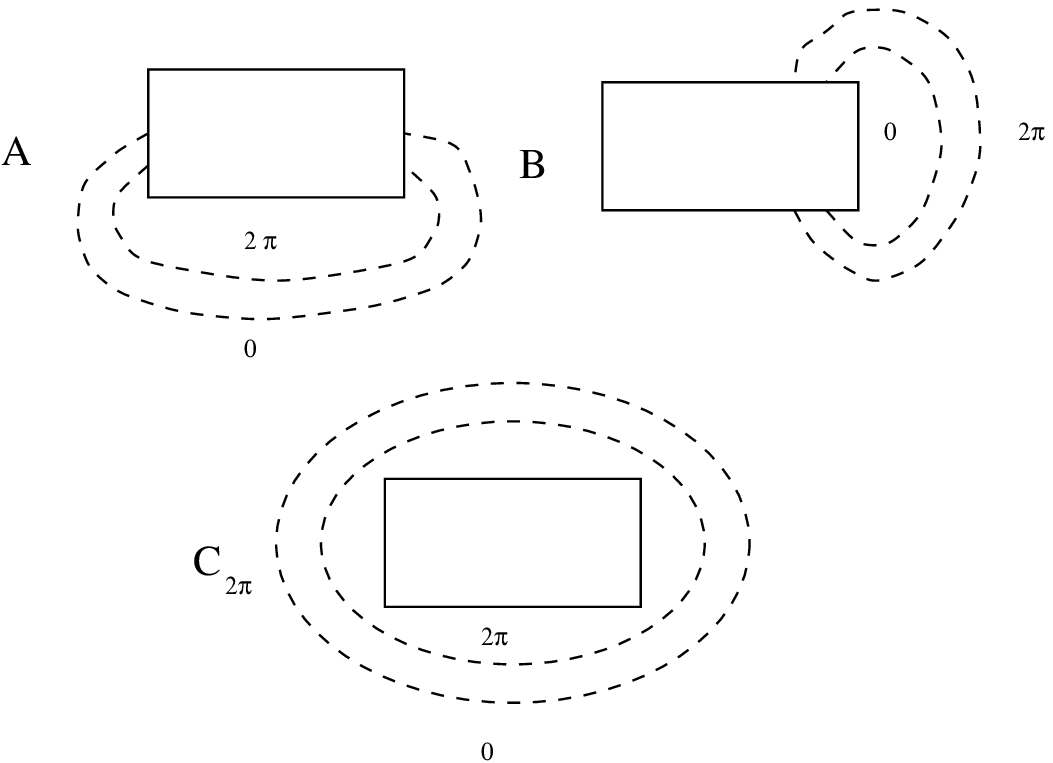}}
{{\bf Fig. 2.2:} Dehn twists corresponding to diffeomorphisms of the mapping
  class group. The annuli enclose loops, which we have
  omitted in the figure. Rotations are counterclockwise by convention.}
\end{figure}

For the present case, $\Sigma $ is the plane with a single geon and the mapping
class group is isomorphic to the central extension of the group
$SL(2,{\bf Z})$, denoted by $St(2,{\bf Z})$ and called the Steinberg
group.  This is the same as the mapping class
group of a torus minus one point \cite{Aneziris}.
We denote generators of $M_{\Sigma }=St(2,{\bf Z}) $ by $A$ and $B$. They 
correspond to (isotopy classes of) diffeomorphisms \footnote{One can
  see from (\ref{mcg}) that the mapping class group consists of isotopy
  classes of diffeomorphisms. Throughout this paper we shall loosely
  use a representative in a class as the class itself.} called Dehn
twists. 
A Dehn twist is realized as follows. Take a loop in $\Sigma$. Then
draw an annulus enclosing the loop and 
introduce radial coordinates $r\in [0,1]$, with $r=0$ and $r=1$ 
corresponding to the boundaries of the annulus, see Fig. 2.2. 
Then rotate the points of the 
annulus in such a way that the angle of rotation $\theta (r)$ is zero
for $r=0$ 
and gradually increases, becoming $2\pi $ at $r=1$. 
Figure 2.2 shows how to produce Dehn twists, and in Fig. 2.3, we show
how the
Dehn twist $B$ deforms the loop $\gamma_1$.
There is also the Dehn twist along a loop enclosing the geon, which
can be interpreted as the $2 \pi$-rotation of the geon
\cite{erice,FS,Aneziris}. It will be important when we discuss
spin of the geon. The corresponding diffeo is denoted by $C_{2\pi}$
in Fig. 2.2. However, $C_{2\pi}$ is not independent of $A$ and $B$.  
One can show that \cite{Aneziris}
\be
C_{2\pi}=(AB^{-1}A)^{4}. 
\ee
The group $M_\Sigma $ is generated by $A$ and $B$, 
with the relation that $C_{2\pi}$ commutes with $A$ and $B$. It is
useful to think of
the elements of $M_\Sigma $ as words $W(A,B)$ in $A$, $B$ and their inverses. 
 
The action of $A$ and $B$ on $[\gamma _i]\in \pi_{1} (\Sigma )$ 
induces an action on $(a,b)\in T$, and therefore induces operators 
$\hat A$ and $\hat B$ in the defining representation acting on
$V$. Let us take as an 
example the action of $B$ on $\gamma _1$, as given by Fig. 2.3. One sees that 
$[\gamma _1] \rt [\gamma _1][\gamma _2]$, and therefore $a\rt ab $. On the 
other hand, $B$ keeps $[\gamma _2]$ invariant.
\begin{figure}[t]
\begin{center}
\begin{picture}(0,0)
\put(80,-7){$P$}
\put(83,3){$\ast$}
\end{picture}
\epsfbox{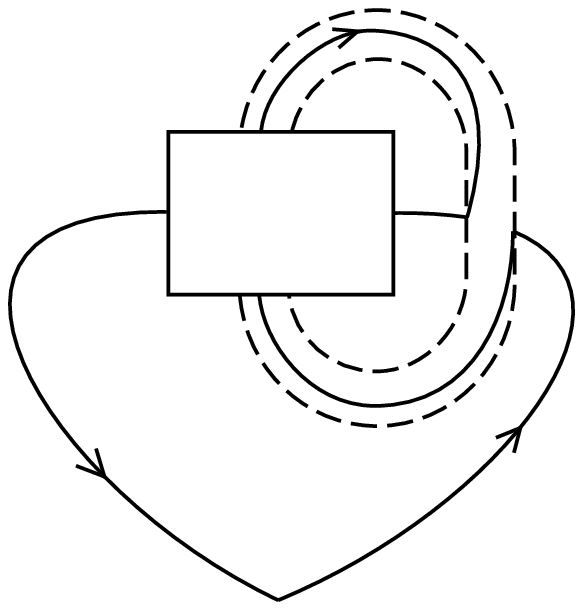}
\end{center}
{{\bf Fig. 2.3:} Dehn twist $B$ and its action on $\gamma_1$.}
\end{figure}
One can verify that $A$ and $B$ induce the following operators:
\bea
\label{abc}
\hat A \mid a,b\rangle \, & =&\, \mid a, ba \rangle \; ,\nonumber\\ 
\hat B \mid a,b\rangle \, &=&\, \mid ab,b\rangle \; ,
\eea
For an arbitrary word $W(A,B)$, the corresponding operator is 
$W(\hat A,\hat B)$, i.e., the
same word but with $A$ and $B$ replaced by $\hat A$ and $\hat B$. 
For example, the Dehn  twist $C_{2\pi}$ of Fig. 2.2 
is written as $(AB^{-1}A)^4$ and the
corresponding operator $\hat C_{2\pi}$ can be immediately computed to be
\be
\label{opC}
\hat C_{2\pi} \mid a,b\rangle \, =\, 
\mid {\bf c}^{-1}a{\bf c} ,{\bf c}^{-1}b{\bf c}\rangle \;,
\ee
where ${\bf c}=aba^{-1}b^{-1}$. 

The algebra generated by the operators $\hat A$ and $\hat B$ is the group 
algebra ${\IC}({\cal M})$.  Together with ${\IC}(H)$ and 
${\cal F}(H\times H)$ it gives us the total algebra ${\cal A}^{(1)}$
for a  single topological geon. 
From the definitions (\ref{projectors}), 
(\ref{del}) and ({\ref{abc}}) one sees that 
\[
\hat\delta _g\hat A=\hat A\hat\delta _g, \,\,\,\,\,\hat\delta _g\hat
B=\hat B\hat\delta _g,
\]
\[
\hat\delta _gQ_{(a,b)}\hat\delta _g^{-1}=Q_{(gag^{-1},gbg^{-1})},
\]
\[
\hat C_{2\pi} \hat A = \hat A \hat C_{2\pi}, \,\,\,\,\, \hat C_{2\pi}
\hat B = \hat B \hat C_{2\pi},
\]
\be
\hat AQ_{(a,b)}\hat A^{-1}=Q_{(a,ba)},\,\,
\hat BQ_{(a,b)}\hat B^{-1}=Q_{(ab,b)}.\label{relations} 
\ee
Therefore, both algebras ${\IC}(H)$ and ${\IC}({\cal M})$ act on 
${\cal F}(H\times H)$.  The action of a generic word
$W(\hat A,\hat B)$ on $Q_{(a,b)}$ will be denoted by
\be
\label{Waction}
W(\hat A,\hat B)Q_{(a,b)}= Q_{(w^{(a)},w^{(b)})}W(\hat A,\hat B).
\ee
where $(w^{(a)},w^{(b)})$ is a pair of words in $a$ and $b$ and their 
inverses, representing the action of $W(A,B)$ on $(a,b)$. 

There are two equivalent ways of presenting ${\cal A}^{(1)}$. One is by using 
the defining representation of (\ref{projectors}), (\ref{del}) and ({\ref{abc}}).
Another way is to define  ${\cal A}^{(1)}$ as the algebra generated by 
$Q_{(a,b)}$, $\hat\delta _g$, $\hat A$ and $\hat B$ with the relations
(\ref{relations}). In any case, we have that 
\be
\label{productinA}
{\cal A}^{(1)} = \IC(H\times{\cal M}) \semidirect {\cal F}(H\times H).
\ee

We may introduce the algebra 
for two topological geons following exactly the same ideas as for a single 
topological geon. We will briefly outline here the main constructions.
For details, see \cite{us}. We recall that for a single geon, ${\cal A}^{(1)}$ 
consists of three sub-algebras, generated by the ``position
observables''  ${\cal F}(T)$, the $H$-transformations 
${\IC}(H)$,  and the ``translations'' , i.e., a realization 
${\cal M}$ of the mapping class group $M_\Sigma $. The algebra 
${\cal A}^{(2)}$ for two geons 
will consist of the same three distinct parts, with 
$T=H\times H\times H \times H \equiv H^4$ and $\Sigma $ replaced by a
plane with two handles.

It is natural to work with the defining representation on $V\otimes V$ 
spanned by vectors of the form
\[
\mid a_1 , b_1 \rangle \otimes \mid a_2 , b_2 \rangle \; ,
\]
where the subscripts denote the respective geons. The ``position
observables'' are generated by projectors $Q_{(a_1 ,b_1 )} \otimes
Q_{(a_2 ,b_2 )}$ acting on $V\otimes V$ in the obvious way, i.e., 
\be
Q_{(a_1 ,b_1 )} \otimes Q_{(a_2 ,b_2 )}
\mid a'_1 , b'_1 \rangle \otimes \mid a'_2 , b'_2 \rangle =
\delta_{a_1,a'_1}\delta_{b_1,b'_1}\delta_{a_2,a'_2}\delta_{b_2,b'_2}
\mid a_1 , b_1 \rangle \otimes \mid a_2 , b_2 \rangle \,.
\label{position}
\ee
Therefore, the  ``position''  operators belong to 
${\cal A}^{(1)}\otimes {\cal A}^{(1)}$.

The action of an $H$-transformation $g\in H$ on the fluxes 
$(a_1,b_1,a_2,b_2)$ is by a global conjugation. This induces the action
\be
\mid a_1 , b_1 \rangle \otimes \mid a_2 , b_2 \rangle \, \rt \
\mid ga_1g^{-1} , gb_1g^{-1} \rangle \otimes 
\mid ga_2g^{-1} , gb_2g^{-1} \rangle  
\ee
on $V\otimes V$. The corresponding operator is obviously identified with  
$\hat \delta _g\otimes \hat \delta _g \in  {\IC}(H)\otimes {\IC}(H)$, since 
\be
\hat \delta _g\otimes \hat \delta _g
\mid a_1 , b_1 \rangle \otimes \mid a_2 , b_2 \rangle=
\mid ga_1g^{-1} , gb_1g^{-1} \rangle \otimes 
\mid ga_2g^{-1} , gb_2g^{-1} \rangle 
\label{2gauge}
\ee
Hence, $H$-transformation operators  also belong to 
${\cal A}^{(1)}\otimes {\cal A}^{(1)}$.

We now start to consider the action of the mapping class group $M_\Sigma $.
For two or more geons, $M_\Sigma $ is much more 
complicated than for a single geon \cite{birman}. The mapping class group
is generated by Dehn twists of the type $A$ and $B$ (see Fig. 2.2) for each 
individual geon together with diffeomorphisms involving pairs of geons.  

Let $A_i,B_i$, $i=1,2$ be the generators of the ``internal diffeos'' for each 
individual geon. The corresponding operators acting on $V\otimes V$ 
are clearly given by
\[
\hat A_1=\hat A\otimes \II, \,\,\,\,\, \hat A_2=\II \otimes \hat A 
\]
\be
\hat B_1=\hat B\otimes \II, \,\,\,\,\, \hat B_2=\II \otimes \hat B
\label{internal}
\ee
where $\II $ is the identity operator on $V$. 

There are two additional classes of 
transformations besides the internal diffeos. The first one, called 
exchange, is the analogue of the elementary braiding of two 
particles. The second, called handle slide, has no analogue for particles,
since it makes use of the internal structure of the geon.

So far, all operators in the algebra for ${\cal A}^{(2)}$ were of the form 
$x\otimes y\in {\cal A}^{(1)}\otimes {\cal A}^{(1)}$. It turns out that 
this is not the case for exchanges and handle slides.
They correspond somewhat to 
interactions and cannot be written strictly in terms of operators in
${\cal A}^{(1)}\otimes {\cal A}^{(1)}$. In order to describe interactions between geons, 
we need to define a pair of flip automorphisms of $V \otimes V$. They
are necessary
in the construction of the exchange and handle slide operators. 

{\bf Definition}: Given a two geon state 
\[
\mid a_1 , b_1 \rangle \otimes \mid a_2 , b_2 \rangle \in V\otimes V\; ,
\]
the flip automorphisms $\sigma $ and $\gamma $ are defined by:
\bea
\sigma \mid a_1 , b_1 \rangle \otimes \mid a_2 , b_2 \rangle \, &:=&\, 
\mid a_2 , b_2 \rangle \otimes \mid a_1 , b_1 \rangle \; , \nonumber\\
\gamma \mid a_1 , b_1 \rangle \otimes \mid a_2 , b_2 \rangle \, &:=&\, 
\mid a_1 , b_2 \rangle \otimes \mid a_2 , b_1 \rangle \; . \nonumber
\eea

Both are not given geometrically as morphisms of the mapping class
group, but unless one introduces these operators, the algebra of two
geons cannot be related directly to the algebras for a single
geon. We will show that the algebra ${\cal A}^{(2)}$ can be obtained from
the tensor product ${\cal A}^{(1)}\otimes {\cal A}^{(1)}$ when we add
$\sigma $ and $\gamma $.

In the exchange process, two geons permute their positions. 
In our convention, the geon
on the right (left) moves counterclockwise to the position of the
left(right) (see
Fig. 2.4).
\begin{figure}
\centerline{\epsfbox{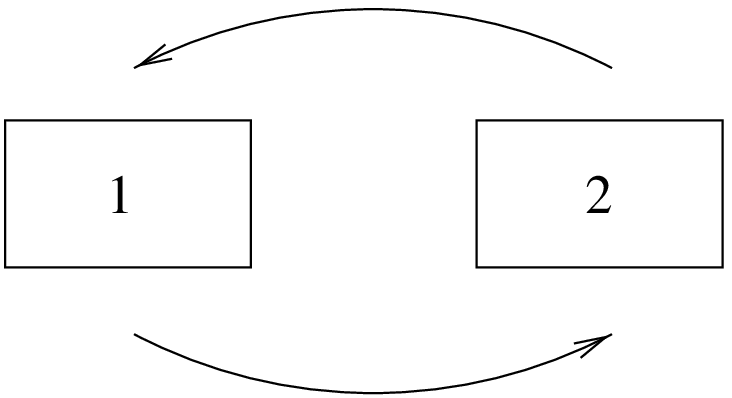}}
\begin{center}
{{\bf Fig. 2.4} Geon exchange.}
\end{center}
\end{figure}
The effect of a geon exchange on the states is of the form
\be
\label{geonexchange}
{\cal R} \mid a_1 , b_1 \rangle \otimes \mid a_2 , b_2 \rangle \,
=\, \mid {\bf c}_{1}^{-1} a_2 {\bf c}_{1},
{\bf c}_{1}^{-1} b_2 {\bf c}_{1}\rangle \otimes 
\mid a_1 , b_1 \rangle \; ,
\ee
where ${\bf c}_{1} = a_1b_1a^{-1}_1b^{-1}_1$. This operator is
equivalent to braiding operators for particles and 
also satisfies the Yang-Baxter equation,
\be
\label{yb}
({\cal R} \otimes {\II })({\II } \otimes {\cal R})
({\cal R} \otimes {\II }) \, =\, ({\II } \otimes {\cal R})
({\cal R} \otimes {\II })(\II \otimes {\cal R}) \; .
\ee
One can verify that the exchange operator (\ref{geonexchange}) 
may be written as the product
\be
\label{R}
{\cal R}=\sigma \, R
\ee
where $R\in {\cal A}^{(1)}\otimes {\cal A}^{(1)}$ is the analogue of the 
universal $R$-matrix for a quasi-triangular Hopf algebra. In our case $R$ is
given  by
\be
\label{R'}
R=\sum _{a,b}Q_{(a,b)}\otimes {\hat \delta}^{-1} _{aba^{-1}b^{-1}} 
\ee

The handle slide ${\cal S}$ is shown in Fig. 2.5. 
In (a), the geon is viewed as a rectangular box on the plane. 
In (b), we have identified two edges of the rectangle and the
geon is represented as two circles on the plane connected by dotted
lines. The action of ${\cal S}$ on the states and its presentation
in terms of the other operators is given in \cite{us}. 

\begin{figure}
\begin{center}
\begin{picture}(0,0)
\end{picture}
\epsfbox{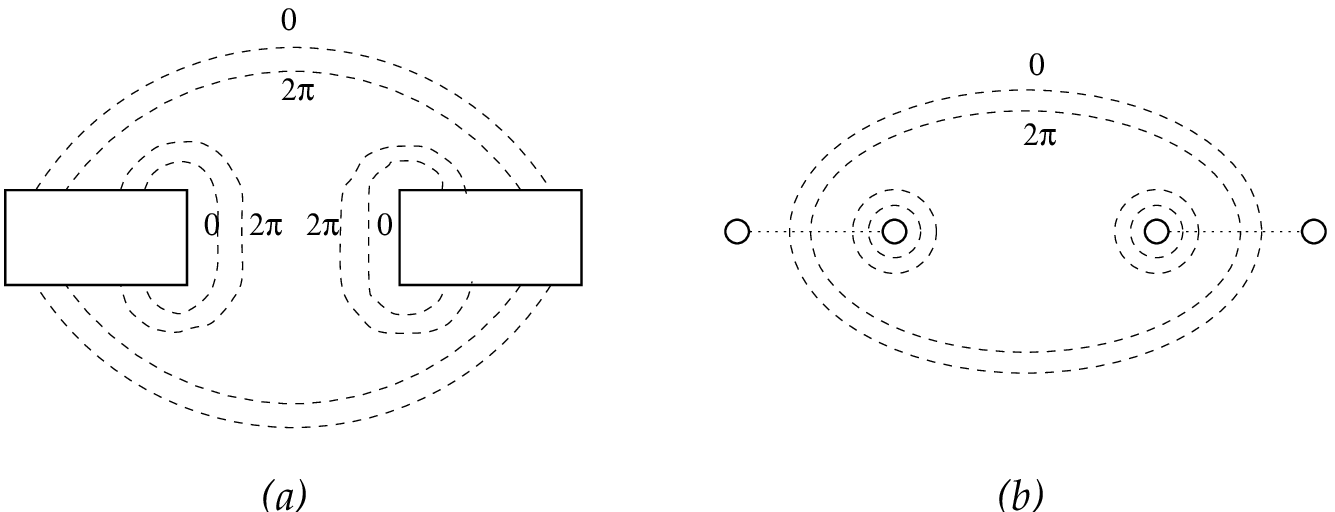}
\end{center}
{{\bf Fig. 2.5:} The handle slide is interpreted geometrically as
  the full monodromy of two handles followed by a rotation of
  $2\pi$ of each handle. The figure shows two equivalent
  representations for the handle slide: In (a), the geon is
  viewed as a rectangular box on the plane. In (b), we have identified
  two edges of the rectangle and the geon is represented as two circles
  on the plane. } 
\end{figure}
This completes the description of ${\cal A}^{(2)}$. The algebra for two geons
is generated by the elements of ${\cal A}^{(1)}\otimes {\cal A}^{(1)}$, 
${\cal R} $ and the handle slide ${\cal  S}$.

These constructions can be easily generalized to write down the algebra
${\cal A}^{(n)}$ for $n$ geons \cite{us}. 

\sxn{Quantization}\label{S4}

The algebra ${\cal A}^{(1)}$ describes the topological degrees of freedom for
a single geon on the plane. To quantize the system we need to find an 
irreducible representation of ${\cal A}^{(1)}$ on a Hilbert space
${\cal H}$. However, this Hilbert space will branch into irreducible
representations of the field algebra:
\be 
{\cal H} = \oplus_{r}{\cal H}_{r},
\ee
where ${\cal H}_{r}$ denotes a particular irreducible representation
describing a certain geon type.
The algebra is finite dimensional, and therefore there will be a
finite number of  
irreducible representations of ${\cal A}^{(1)}$. Furthermore, the Hilbert 
spaces ${\cal H}$ are all finite dimensional. Each representation gives us a
possible one-geon sector of the theory.

In the case of quantum doubles, the irreducible representations
are fully classified. See for 
instance ref. \cite{SV}. For the case of geons, the algebra is more
complicated because of the existence of internal
structure. Nevertheless, the representations of ${\cal A}^{(1)}$ are quite 
similar to the ones of the quantum double of a finite group. This is
not totally surprising,
since in a certain limit, as discussed in the previous
section, we recover the quantum double ${\cal D}^{(1)}\cong D(H)$. Actually,
we can define a class of  algebras ${\cal A}$, called transformation
group algebras, that can have its representations
classified and that are generic enough to contain the quantum double 
and the algebra ${\cal A}^{(1)}$ as particular cases.
In the spirit of \cite{SV}, one can then get all representations  of
${\cal A}$. 

\noindent
{\bf Definition}: Let $X$ be a finite set and $G$ a finite group acting on 
$X$. In other words, there is a map $\alpha _g:X\rt X$ for each $g\in G$.  
As usual, we denote by ${\cal F}(X)$ the algebra of functions on $X$ and 
by $\IC (G)$ the group algebra of $G$.  
We define the algebra ${\cal A}$ as the vector space 
\[
{\cal A}:={\cal F}(X)\otimes \IC (G)
\]
with basis elements denoted by $(Q_{x},g)$, $Q_{x} \in {\cal F}(X)$ and $g
\in \IC (G)$, and the multiplication 
\be
\label{prodinA}
(Q_x , g ) \cdot (Q_y ,h )\,:=\, 
(Q_x Q_{\alpha_{g} (y)} , g h ) \; .
\ee
Here, $Q_x$ is the characteristic function supported at $x \in X$. Let
$x_0$ be an element of $X$. We denote by  $K_{x_0}\subset G$ the 
stability subgroup with respect to $x_0$, i.e.,
\be
K_{x_0} \, =\, \left\{ g\in G \mid  \alpha_g (x_0 )=x_0 \right\} \; .
\ee

The stability subgroup $K_{x_0}$ divides  the group $G$
into equivalence classes of left cosets. Let $N$ be the number of
equivalence classes and let us choose 
a representative $\xi_i \in G$, $i=1\ldots N$ for each class,  with the 
convention that  $\xi_1 =e$. We can write the following partition of
$G$ into left cosets: 
\be
G\, =\, \xi_1 K_{x_0} \cup \xi_2 K_{x_0} \cup \ldots \cup \xi_N K_{x_0} \; .
\ee

We can now state the following result.   

\noindent
{\bf Theorem }
Let $\mid j \rangle_{\rho}$, $j=1\ldots n$ be a basis of a subspace $V_\rho$ of
$\IC(G)$ carrying an IRR $\rho$ of $K_{x_0}$. Then, for (a fixed) $x_0 \in X$, 
elements $\xi_i \in G$, $i=1\ldots N$ and $\mid j \rangle_{\rho} \in
{\IC} (G)$, $j=1\ldots n$ as stated above, the vectors
\[
\xi_i \mid x_0 \rangle \otimes  \mid j \rangle_{\rho} :=
\mid \alpha_{\xi_i}(x_0) \rangle \otimes  \mid j \rangle_{\rho} \; ,
\]
form a basis for an IRR of the algebra 
${\cal A}$, given by 
\[
(Q_x , g )\mid \alpha_{\xi_i}(x_0) \rangle \otimes  \mid j
\rangle_{\rho} := \delta_{x, \alpha _{\xi_{i'}}(x_0)}\mid \alpha_{\xi_{i'}}(x_0)
\rangle \otimes\Gamma^{(\rho)}(\beta)_{kj} \mid k \rangle_{\rho}\; ,
\]
where $\xi_{i'}$ and $\beta$ are uniquely determined by the equation 
\[
g\xi_i = \xi_{i'}\beta \; ,
\]
and $\Gamma^{(\rho)}$ is the matrix for the representation $\rho$.  

This result follows from a standard construction in induced
representation theory (cf. discussion of the Poincar\'e group in
\cite{representation}). 

The quantum double $D(H)$ and the algebra ${\cal A}^{(1)}$ are 
particular cases of transformation group algebras. The quantum double
is obtained by 
taking $X=H$, $G=H$, with the action $\alpha _g(h)=ghg^{-1}$. As for the
algebra of a single geon, one takes
\[
X=H\times H\,\,\,
\] 
and for the group $G$ the product $H\times {\cal M}$. The actions of 
$\hat\delta _g\in H$ and
$W\in {\cal M}$ commute and are given by 
\[
\alpha_g (a,b) \, =\, 
(gag^{-1} ,gbg^{-1}), \;\,\,\,g\in H 
\]
and 
\[
\alpha_W (a,b) \, =\, (w^{(a)},w^{(b)}), \;\,\,\, W\in {\cal M} ,
\]
where we have used the notation of (\ref{Waction}). 
The IRR's for the algebra (\ref{productinA}) can be
constructed given an element $(a,b) \in H\times H$. The stability
subgroup $K_{(a,b)}\subset H \times H$ is defined by 
\be
K_{(a,b)} \, =\, \left\{ (g,W) \in H\times {\cal M} \mid 
\alpha_g\alpha_W (a,b):=(gw^{(a)}g^{-1},gw^{(b)}g^{-1})=(a,b) \right\} \; . 
\ee
Then, after choosing representatives $\xi_1 , \ldots , \xi_N$ for the 
left cosets, the partition of $H\times {\cal M}$ can be written as
\be
\label{K}
H\times {\cal M} \, =\, \xi_1 K_{(a,b)} \cup \xi_2 K_{(a,b)} \cup \cdots \cup 
\xi_N K_{(a,b)} \; .
\ee
Let $\mid 1 \rangle, \ldots , \mid n \rangle \in {\IC} (H\times {\cal
  M})$ be a basis of an IRR of $K_{(a,b)}$.
Then, according to the theorem, the vectors
\be
\label{IREP}
\mid \alpha_{\xi_i}(a, b) \rangle \otimes  \mid j \rangle_{\rho} \;,
\ee
with $i=1\ldots N$, $j=1\ldots n$,
form a basis of an IRR of the algebra 
${\cal A}^{(1)}$. 

Let us express the representations of ${\cal A}^{(1)}$ in a more compact 
notation. 
The action of $H\times {\cal M}$ on $X=H\times H$ divides $X$
into orbits. We denote by $[a,b]$ the orbit containing  the element 
$(a,b)\in H\times H$. We will collectivelly call $\rho $ the quantum numbers  
labeling  the IRR's  of  $K_{(a,b)}$. 
One can see from (\ref{IREP}) that an IRR  $r$ is 
characterized by a pair $r=([a,b],\rho )$. A basis for an IRR $r$ 
of ${\cal A}^{(1)}$ will therefore be written as vectors 
$\mid i,j\rangle ^{(a,b)}_r$ ,$i=1,...,N$; $j=1,...,n$ defined by
\be
\label{IREPvec}
\mid i,j\rangle ^{(a,b)}_r:=\xi_i \mid a,b\rangle\otimes 
\mid j\rangle _\rho   
\ee
where $\mid a,b\rangle $ is a state in the defining representation, 
$\xi_i$ are the same as in (\ref{K}) and $\mid j\rangle _\rho $ are  
base elements in the irreducible representations $\rho $ of
$K_{(a,b)}$. Of course, the set of vectors thus defined depend on the
pair $(a,b)$ we choose. We fix an $a$ and a $b$, and henceforth omit
the superscript. 

The action of $Q_{(a',b')}$ is given by
\be
Q_{(a',b')}|i,j>_r=Q_{(a',b')}~~\xi_i|a,b>\otimes |j>_\rho=
Q_{(a',b')}|a_i,b_i>\otimes |j>_\rho=\delta_{a',a_i}\delta_{b',b_i}|i,j>_r.
\ee

Let $\hat\delta_gW$ be a generic element of  $H\times {\cal M}$. The equation
\be
\label{gaction}
\hat\delta_gW\,\,\xi_i=\xi_{i'}\beta 
\ee
defines uniquely a new class $\xi_{i'}$, together with an element of the 
stability group $\beta \in K_{(a,b)}$. 
The action  of $\hat \delta _gW\in {\cal A}^{(1)}$ on
$\mid i,j\rangle _r$ is determined by (\ref{gaction}) and it reads
\bea
\label{Aaction}
\hat \delta _gW\mid i,j\rangle _r &=& 
\xi_{i'}\mid a,b\rangle\otimes 
\beta \mid j\rangle _\rho =\nonumber \\
&=& \sum_{k}\,{\Gamma ^{(\rho)}(\beta )}_{kj}
\mid i',k\rangle _r
\eea
where $\Gamma ^{(\rho )}$ is the matrix representation of $K_{(a,b)}$.

Each IRR $r=([a,b],\rho )$ describes a distinct 
quantum geon. The corresponding vector spaces ${\cal H}_r$ generated by states
$\mid i,j\rangle _r $, are all finite dimensional. Therefore we can 
easily make it into a Hilbert space by introducing the scalar product
\be
\langle i',j'\mid i,j \rangle _r=\delta _{ii'}\delta _{jj'}.
\ee  

Since the algebras ${\cal A}^{(1)}$ are not the same for different 
choices of the discrete group $H$, we cannot say in general what is
the spectrum of a geon. First, we need to fix a group $H$ and then 
compute the spectrum for the corresponding ${\cal A}^{(1)}$.

Consider now two geons described by representations $r_1$ and $r_2$. The 
associated  Hilbert space of states is simply 
\be
\label{2geonsHilbert}
{\cal H}^{(12)}:={\cal H}_{r_1}\otimes {\cal H}_{r_2}.
\ee
As explained in Section \ref{S3}, the field algebra consists of 
${\cal A}^{(1)}\otimes {\cal A}^{(1)}$ together with ${\cal R}$
and  ${\cal S}$. The elements of ${\cal A}^{(1)}\otimes {\cal A}^{(1)}$ act
naturally on (\ref{2geonsHilbert}). It remains to be said what is the
action of  ${\cal R}$ and ${\cal S}$ on states in ${\cal
  H}_{r_1}\otimes {\cal H}_{r_2}$.  

The action of ${\cal R}$ is completely determined by the formula
(\ref{R}):
\[
{\cal R}=\sigma \sum _{a,b}Q_{(a,b)}\otimes {\hat \delta}^{-1} _{aba^{-1}b^{-1}}.
\]
In other words
\be
\label{actR}
{\cal R}\mid i,j \rangle _{r_1}\otimes \mid k,l\rangle _{r_2}=
\sum _{a,b}{\hat \delta}^{-1}_{aba^{-1}b^{-1}}\mid k,l \rangle _{r_2}\otimes 
Q_{(a,b)} \mid i,j\rangle _{r_1}
\ee.

The generalization for $n$ geons is straightforward.

We may think of ${\cal R}$ and ${\cal S}$ as scattering matrices for a
pair of geons. The ${\cal R}$-matrix represents an ``elastic'' interaction in
the sense that two incoming geons of quantum numbers $r_1$ and $r_2$
are scattered into two objects carrying the same quantum numbers 
$r_1$ and $r_2$. The
handle slide ${\cal S}$ on the contrary is a nontrivial
scattering, each one of the two outgoing geons being a
superposition of many geons in the spectrum. 

\sxn{The Spin-Statistics Connection for $(2+1)d$ Topological Geons}\label{S6}

The spin-statistics theorem is a well-established
law of physics, and it holds true for most of the quantum particles in
nature. In considering quantum
topology change and, more generally, quantum gravity, one is naturally
led to inquire whether a sort of
field theory exists which gives geons as quantum excitations on a
topologically trivial classical background. Such a theory remains
hitherto utterly elusive, but one may try to investigate some of its
aspects. For instance, in such a theory the quanta would be geons, so
one fundamental research work would be a thorough analysis of the spin
and statistics
of geons and in particular whether the geons enjoy the canonical
spin-statistics connection. 

In order to pose the problem properly, we start with some general
comments. Spin and statistics are two properties which can be defined
independently of each other. They refer not only to particles, but in
general to localized sub-systems that may consist of  several
particles or even extended objects like solitons. The tensorial or
spinorial nature of spin
is determined by the behavior of the wave function when the sub-system
undergoes a ${2\pi }$ rotation. 
Statistics determines what happens
when two identical sub-systems are exchanged. The canonical
spin-statistics connection asserts the identity of the operators
implementing $2\pi$-rotation and exchange. It is thus clear that our
first task is to define the quantum operators responsible for ${2\pi
  }$ rotation and exchange. In the case of $(2+1)d$ orientable geons,
we can easily find these operators among the algebra
described in Section \ref{S3}. In doing so, we will be led to a
definite spin-statistics relation differing from the canonical one by
a phase. 

In the usual spin-statistics theorem, we prepare a two-particle state
which is a tensor product of two copies of the same (arbitrary) one-particle
state with the same spin, in order to probe the relation between the
exchange and $2\pi$-rotation operators. However, here we meet a
problem: we have only been able to prove the existence of such
a relation in the special case when the total fluxes of each geon
{\em not only belong to the same conjugation class, but are the
  same} . This is rather unexpected at first sight, because in the
definition of the IRR's of the one-geon algebra, eq. (\ref{IREPvec}),
we can see that each vector of the basis of a representation space
has an associated and possibly distinct flux $c_i 
= \xi_i c \xi_i^{-1}$, where $c = aba^{-1}b^{-1}$. The
geon state is in general a superposition of such vectors, and
therefore the flux of a geon state is not well defined for most
states in a given representation. To overcome this problem we advance
the claim {\em that 
the total flux should be a superselected quantity in the same sense
that electric charge is superselected}. We
shall explain how this superselection occurs in Section
\ref{Newsection}. Here we
simply note that as a consequence, the only {\em physical} states are
superpositions of basis vectors with the same total flux. If we assume
this, then the spin-statistics connection holds for all {\em physical} states.

We start by considering a pair of geons on the plane. The algebra 
${\cal A}^{(2)}$ is fixed as soon as we choose a finite group $H$. In Section 
\ref{S3} we saw that the exchange of two geons is realized by the operator
${\cal R}$ defined in (\ref{R}). Similarly, the $2\pi $ rotation of one
of the geons can be written 
as $\hat C_{2\pi }\otimes \II$ or $\II \otimes \hat C_{2\pi }$. We would like 
to know if the algebra ${\cal A}^{(2)}$ implies any kind of relation 
between exchange and $2\pi $ rotations. 

Let us start by fixing a representation $r=([a,b],\rho )$ of ${\cal
  A}^{(1)}$. State vectors $\mid i,j\rangle _{r}\in {\cal H}_r$ given by
  (\ref{IREPvec}) do not have a well-defined
spin, or in other words, they are not eigenstates of $\hat
C_{2\pi}$. In order to establish the spin-statistics relation we need
first to know what are the eigenstates of $\hat C_{2\pi }$. We recall
that $\hat C_{2\pi }\in {\cal M}$ and that ${\cal M}$ is a finite
group. Therefore there exists a least non-negative integer $N$ such that 
\be
{\hat C_{2\pi}}^N=1
\ee
and the eigenvalues of $\hat C_{2\pi }$ are $e^{i2\pi s}$, where the
spin $s=0,1/N,2/N,...,(N-1)/N$. One can see that the operator
\be
\label{spin1}
P_s= \frac{1}{N} \sum _{n=0}^{N-1}e^{-in2\pi s}\hat C_{2\pi}^n
\ee
projects into states with spin $s$. The front factor is just a
normalization constant to ensure that $P_{s}^{2} = P_s$, and we assume
of course that $N \geq 2$, so that we have a non-trivial behavior
under $2\pi$-rotation. Furthermore, any state of ${\cal
  H}_r$ with spin
$s$ can be obtained by linear combinations of 
$\mid s;~ij\rangle $ given by
\be
\label{basis}
\mid s;~i,j\rangle =P_s\mid ij\rangle _r. 
\ee

Consider a pair of geons both carrying the same representation 
$r=([a,b],\rho )$
and the same spin $s$. In addition, let us consider both geons to have
the same total flux $c_i = \xi_i c \xi_i^{-1}$, where
$c=aba^{-1}b^{-1}$, i.e., we assume that each geon is in a state which
is the superposition of basis vectors having the same flux. This will
have the following
consequence. The $\xi_i$'s in $(\ref{IREP})$ act on $\mid a,b
\rangle$ in such a way that the flux $c$ becomes conjugated by
$\xi_i$. In other words,  the flux of each vector in the basis
$\{\mid i,j\rangle _{r} \}$ will be equal to $c_i$. Hence $c_i$ becomes a
quantity characterizing the particular superselection sector within $r$.We are
going to show that the statistics of the system is 
completely determined by the spin $s$ and the representation $r$. 
It does not depend on $c_{i}$. More precisely, 
\be
\label{ssc}
{\cal R}\mid \psi \rangle \otimes \mid \phi \rangle =e^{i
  (2\pi s - \theta _r)}\mid \phi \rangle \otimes \mid \psi \rangle,
\ee
where the states $\mid \psi \rangle$ and $\mid \phi \rangle$ transform
  according to the same represention $r$ and have both the same spin
  $s$ and the same total flux $c_i$. Also, $\theta _r$ is an angle that depends
  only on the representation $r$.
The usual spin-statistics connection is true only for 
representations $r$ such that $\theta _r=0$. 

It is enough to show that (\ref{ssc}) is true for states $\mid \psi
\rangle$ and $\mid \phi \rangle$ in the basis (\ref{basis}):
\bea
\label{basis2}
\mid \psi \rangle =P_s\mid ij\rangle _r &=&
\frac{1}{N} \sum _{n=0}^{N-1}e^{-i n2\pi s}\hat C_{2\pi }^n
\{\ \xi_i \mid a,b\rangle\otimes \mid j\rangle _\rho\}, \nonumber \\
\mid \phi \rangle =P_s\mid kl\rangle _r &=&
\frac{1}{N} \sum _{n=0}^{N-1}e^{-i n2\pi s}\hat C_{2\pi }^n
\{\ \xi_k \mid a,b\rangle\otimes \mid l\rangle _\rho\}, \nonumber \\
c_i &=& c_k.
\eea
Using (\ref{actR}) one can show after some algebra that 
\be
\label{5}
{\cal R}~\{P_s\mid ij\rangle _r \otimes P_s\mid kl\rangle _r \} =
\mid \Phi \rangle \otimes P_s\mid ij \rangle _r .
\ee
with
\be
\mid \Phi \rangle =\frac{1}{N} \sum_n e^{-in2\pi s}\hat C_{2\pi}^n~~
\hat\delta_{c_i}^{-1}~\{\ \mid a_k,b_k \rangle \otimes \mid l\rangle
_\rho \}.
\ee
In the last formula, we used the notation
\be
\mid a_k,b_k\rangle:=\xi _k\mid a,b\rangle , 
\ee 
with $c_i$, the total flux of $\mid a_i,b_i\rangle$, being given by
$c_i=a_ib_ia_i^{-1}b_i^{-1}$. The result (~\ref{5}) is a tensor
product where only the first factor is not yet in the desirable
form of (a phase)$\times \mid \psi \rangle$.  

Let us take a closer look at $\mid \Phi \rangle $. First let us define 
\be
E_i=\hat \delta _{c_i}\hat C_{2\pi }
\ee
and write $\mid \Phi \rangle $ as
\be
\label{6}
\mid \Phi \rangle=e^{i 2\pi s}\frac{1}{N} \sum_n e^{-in2\pi s}\hat C_{2\pi}^nE^{-1}_i
\{\ \xi_k \mid a,b\rangle
\otimes \mid l\rangle _\rho\}.
\ee

To proceed, we will show that 
\be
E_i\xi_k=\xi_k E_1  \label{confusing},
\ee 
where $E_1=\hat \delta _{c}\hat C_{2\pi }$ with
$c=aba^{-1}b^{-1}$. Indeed, first note that
\be
E_i\xi_k|a,b>=\hat\delta _{c_i}C_{2\pi }\xi_k|a,b>=
\hat\delta _{c_i}C_{2\pi } |a_k,b_k>=|a_k,b_k>=\xi_k|a,b>,
\ee
where we have used the fact that $c_i = c_k$. Therefore
\be
E_i\xi_k=\xi_k~\beta \label{1},
\ee
where $\beta $ belongs to the stability group $K_{(a,b)}$ of $|a,b>$.

All we have to do is to show that in fact $\beta =E_1:=\hat\delta
_{c}C_{2\pi }$. We recall that by definition $\xi_1=e$.

From (\ref{1}) it follows that
\be
\beta = \xi_k^{-1}E_i\xi_k.
\ee
If we take a generic $\xi_k$ to be of the form $\hat\delta_{g_k}\omega _k$
for some $g_k\in H$ and $\omega _k\in {\cal M}$, we see that
\be
\beta =  \omega_k^{-1} \hat\delta^{-1}_{g_k}(~\hat\delta _{c_i}C_{2\pi }~)
\delta_{g_k}\omega_k= \hat\delta^{-1}_{g_k}(~\hat\delta _{c_i}C_{2\pi
}~)\hat\delta_{g_k}=\hat\delta_{g_k^{-1}c_ig_k}C_{2\pi },\label{2}
\ee
since $\omega _k$ commutes with $H$-transformations. We will show
presently that 
\be
c_i = c _k = g_kcg_k^{-1},\label{3}
\ee 
which allows us to conclude, from (\ref{2}), that  $\beta =E_1$, and
hence (\ref{confusing}) follows. In
order to show (\ref{3}) we write
\be
|a_k,b_k>=\xi_k|a,b>:=\hat\delta_{g_k}\omega_k|a,b>=
\omega_k|g_kag_k^{-1},g_kbg_k^{-1}>.
\ee
Remembering that the elements of ${\cal M}$ do not change the total flux,
we can compare the total fluxes of the first and last expressions. That
gives us
\be
c_k=g_kcg_k^{-1}.
\ee

Now, from (\ref{confusing}) we have
\be
E_i^{-1}\xi_k=\xi_kE_1^{-1}.
\ee
It is a simple exercise to show that $E_1$ is in the center
of the stability group $K_{(a,b)}$ of $\mid a,b\rangle $. Therefore,
from (\ref{Aaction}), we can write 
\be
E^{-1}_i\{\xi_k \mid a,b\rangle \otimes \mid l\rangle _\rho\}=
\xi_k \mid a,b\rangle \otimes E^{-1}_1\mid l\rangle _\rho\  \label{comute}.
\ee
Since the
states $\mid l\rangle _\rho $ form a basis for an (unitary) IRR of $K_{(a,b)}$,
from Schur's lemma we conclude that the operator $E_1$ gives a phase
$e^{i\theta _r}$ in (~\ref{comute}) depending on the IRR $r=([a,b],\rho
)$. Putting the
results together we have 
\be
\mid \Phi \rangle= 
e^{i (2\pi s -\theta _r)} \frac{1}{N} \sum_n e^{-in2\pi s}\hat C_{2\pi}^n
\{\xi_k \mid a,b\rangle \otimes \mid l\rangle _\rho\} =
e^{i (2\pi s -\theta _r)}P_s \mid kl \rangle .
\ee 
This concludes the demonstration of equation (\ref{ssc}). Since the
phase $\theta _r$ comes from general considerations, it is
indeterminate in our formalism.

\sxn{The Role of $H$: Superselection and Clustering}\label{Newsection}

In this short Section we are interested in emphasizing how the hypothesis of
superselection of total fluxes and the possibity of clustering of geon
subsystems arise naturally if we confine ourselves to local operators
as the only ones with physical significance.

We therefore start off by noting that the $H$-transformations in our
one-geon algebra are actually non-local operations, since they are
effectively  equivalent, from the standpoint of the one geon, to
move ``distant geons'', or alternatively, ``distant fluxes'' in a
circle at infinity around the geon. This apparently innocent
observation entails a striking conclusion: since all elements of the
algebra other than the $H$-transformations (which elements are
also ``local
operators''), commute with the total flux, we have that {\em the total flux
must be a superselected quantity}. As we saw in the previous Section,
this is enough to ensure the Spin-Statistics connection for
geons. This may appear to be strange at first: we have seen that the
total flux {\em can} be changed by the action of $H$. However, since 
$H$-transformations are non-local, they cannot affect local observables in a
sensible, local theory. The only local operators would therefore be
the elements of the mapping class group, and these clearly do not
change the total flux. This can be generalized for $N$ geons: the
total flux of the entire $N$-geon system is then superselected, the
flux of each geon being however subject to changes, much as it occurs
in $QED$ charged sectors. However, in this case one must have a
suitable notion of {\em clustering}, i.e., if we consider $N - 1$
geons to be ``distant'' in a suitable sense, the remaining
``solitary'' geon can be seen as really isolated, and it must be in a
state belonging to a one-geon algebra IRR. We will explain the
necessity of these concepts in what follows.   

First of all we must make precise what our notion of ``distant'' is. We
should recall that we are actually studying the low-energy limit of
a field theory, and in spite of the fact that the theory in this limit
is effectively independent of the metric, we have started off with a
metric in order to define our theory. Therefore it makes sense to
use this metric to measure distances and assume ``distant'' geons or
fluxes to be those which are much further than the characteristic size
of the geon under consideration. We hereafter take this to be the
meaning of ``distant''. ``Local'' will then mean within distances
comparable to the characteristic size of the geon. 

With these definitions we can now proceed to clarify how the notions of
clustering and superselection should arise to ensure the
spin-statistics connection for geons. To establish this connection in
the previous Section we considered states of the form $\mid \psi
\rangle \otimes \mid \phi \rangle $, where $\mid \psi
\rangle $ and $\mid \phi \rangle $ were states transforming in the
same IRR $r$ and having the same spin $s$. We made, in the course of
our demonstration, two basic asumptions:

\begin{itemize}

\item The states  $\mid \psi \rangle $ and $\mid \phi
\rangle $ were considered to have definite, well-defined total fluxes $c_i$
and $c_k$, i.e., they were each superpositions of basis vectors having the same
total flux;

\item The fluxes $c_i$ and $c_k$ are the same.

\end{itemize}

The second asumption is very natural and easily implemented, since as
pointed out in \cite{preskill} in a similar context, the fluxes
``being equal'' is a
well-defined, $H$-invariant notion. One only has to prepare two
one-geon states with same total flux. The first asumption, however,
looks quite artificial at first sight, without some extra input. After
all, each IRR
vector space is generated by vectors having different total fluxes,
although in the same conjugacy class. To impose that only
superpositions of vectors having the same flux should be considered apparently
threatens generality. This is not so however, if we assume that the
notions of clustering of geon subsystems and superselection of total
flux should play a role in our physical description. First, let us
consider the clustering property. Consider a system of $N$ geons on
the plane. It is really rather natural to
assume that geon subsystems can be isolated as long as we consider
only local operators. This means that, if we fix one geon and take the
other $N - 1$ geons to infinity, this remaining geon should be described by the
one-geon algebra ${\cal A}^{(1)}$, and its state should belong to an
IRR of  ${\cal A}^{(1)}$. On the other hand, the total flux of the
$N$-geon system is superselected, but the total fluxes of individual geons
are not uniquely defined by the $N$-geon total flux, and in particular
a geon can even be in a superposition of various
flux states. It is to preclude this possibility that the clustering
property appears: since the geon can be isolated, it is effectively
equivalent to a one-geon system, and in particular all operations
which change its flux are indistinguishable from $H$-transformations
on a one-geon system. Therefore {\em the isolated geon in a pure state
  must have a definite total flux as well}. (Impure one-geon states
with a probability distribution of total fluxes are of course permitted).

The consequence of the previous  discussion for a two-geon system is
as follows. For a
one-geon system, superselection of the total flux means that the
IRR's $r$ must further split into superselected subspaces, and the
physical states belong to these subspaces. Now, for a two-geon system
in a state  $\mid \psi \rangle \otimes \mid \phi \rangle $, only the
total flux must be superselected, but the clustering property ensures
that {\em each} geon must have a definite flux. Therefore for physical
states,  $\mid \psi \rangle $ and $\mid \phi \rangle $ must separately
be superpositions of states with definite flux. The conclusion is that
the two asumptions we made to check the spin-statistics connections
become most natural if we include superselection of the total flux and
the clustering property as physical requirements of the theory.

\sxn{Geons in Quantum Gravity}\label{S7}

In this Section our aim is to describe what are the consequences
of our results to quantum gravity. In the canonical metric formalism of
gravity on a spacetime manifold ${\cal M}$ of the form $\Sigma \times
\IR$, we may perform a space versus time splitting and define the classical
configuration space $Q$ as follows. Let us specialize to $(2 +
1)d$. Let $R^{\infty}(\Sigma)$ be the
space of all Riemannian metrics on 
the space manifold $\Sigma$ which are equal to some
{\em conical} metric in a neighborhood ${\cal N}$ of
infinity\footnote[1]{These boundary conditions  for the metric substitute the
  usual $(3 + 1)d$ ``asymptotically Minkowskian'' scenario for the $(2
  + 1)d$ case. See \cite{Sam} and references therein for a more
  complete discussion.}. We quotient this space by the group
$Diff^{\infty}(\Sigma)$ of diffeomorphisms of the spatial
2-manifold $\Sigma$ to obtain $Q$ \cite{balachandran}. Therefore we
may denote the $Q$ as
\be
\label{QG1}
Q = \frac{R^{\infty}(\Sigma)}{Diff^{\infty}(\Sigma)} \,,
\ee

Now, $Q$ is not simply-connected in general, and in this case one may
have many distinct quantizations, a phenomenon which is seen for
example in the ``$\theta$ vacua'' of QCD\footnote[2]{In this section, the word 
  ``quantization'' will have a meaning slightly different from the rest of
  the paper. It will mean simply an appropriate assignment of a
  Hilbert space of 
  wave functions to a classical configuration space.}. To see this we
resort to the so-called covering space quantization, which we briefly
review here (for a more complete account, see
Ref. \cite{balachandran}). In this approach, wave functions (defining
domains of operators like the Hamiltonian) are taken
to be functions on the universal covering space $\tilde Q$ of $Q$ with
certain specific properties. It
can be shown that in our case 
\be 
\label{QG2}
\tilde Q =   \frac{R^{\infty}(\Sigma)}{Diff^{\infty}_{0}(\Sigma)} \,,
\ee
where $Diff^{\infty}_{0}(\Sigma)$ is the component connected to the
identity, and hence a normal subgroup, of $Diff^{\infty}(\Sigma)$. In
what follows it will always be understood that these diffeos are of
$\Sigma$, and hence we will omit the argument.

It is well known that the universal covering $\tilde X$ of any
topological space $X$ is a principal bundle over $X$ with structure
group $\pi_1(X)$, and the projection $p:{\tilde X }\rightarrow X$ given
by the covering map. The structure group has a free right action
$\gamma:{\tilde x} \rightarrow {\tilde x}\gamma$ on
$\tilde X$ and the quotient ${\tilde X}/\pi_1(X)$ by this action is
again $X$. This action is fiber-preserving, i.e., for every $\gamma
\in \pi_1(X), {\tilde x} \in {\tilde X}$ we have $p({\tilde x}\gamma)
= p(\tilde x)$. Therefore $\tilde Q$ is a principal bundle over
$Q$ with structure group $\pi_1(Q)$, which can inferred from (\ref{QG1})
and (\ref{QG2}) to be isomorphic to the mapping class group
$M_{\Sigma}$. 

Wave functions $\Psi$ need not be single-valued on $Q$ if $Q$ contains
non-contractible loops. Actually the transformation of the wave
function when such loops are traversed gives a representation of
$\pi_1(Q)$. This can be seen in the following way. All that must be
single-valued are observables like
$\Psi^{\ast}\Psi$. On the other hand, since the universal covering space is by
definition simply-connected, functions on it can always be taken to be
single-valued. Therefore, if we could define wave functions as
functions on $\tilde Q$ such that $\Psi^{\ast}\Psi$ is still a function on
$Q$, we would circumvent the multi-valuedness of the wave functions
while leaving the probability interpretation unharmed. This task is
easily accomplished in the following way. Let ${\cal H}_{\rho}$ be a complex
vector space with a hermitian inner product $\langle~ ,~ \rangle$
carrying a unitary representation $\rho$ of $\pi_1(Q)
\simeq M_{\Sigma}$. We say that a function $\Psi :{\tilde Q}
\rightarrow {\cal H}_{\rho}$ is {\em equivariant} if for every $\gamma \in
\pi_1(Q),{\tilde q} \in \tilde Q$ we have 
\be
\label{QG3}
\Psi({\tilde q} \gamma) = \rho(\gamma^{-1})\Psi({\tilde q}).
\ee
Hence, since ${\tilde Q}/\pi_1(Q) = Q$, we have that $\langle
\Psi,\Psi \rangle$ can indeed be viewed as a function on $Q$, as we
wanted. Therefore our quantum Hilbert space $V_{\rho}$ may be taken to
be (the norm completion of) the space of equivariant functions $\Psi
:{\tilde Q} \rightarrow
{\cal H}_{\rho}$ such that the function $\langle \Psi,\Psi \rangle$, seen as
a function on $Q$, gives a finite number when properly integrated on
the whole of $Q$ (the latter process defines the inner product). This
recipe obviously depends on which
representation $\rho$ we take. Actually, it can be shown
\cite{balachandran} that there are at least as many inequivalent
quantizations as there are unitary irreducible representations of $M_{\Sigma}$. 
 
As we have mentioned, wave functions on $Q$ pick an
``Aharonov-Bohm phase'' whenever they traverse a non-contractible path
in $Q$, i.e., they transform according to some representation of
$\pi_1(Q)$. Since this group is non-abelian in general, the wave
function may transform via a non-abelian representation. This is akin
to the behaviour of sections of a
vector bundle with a flat connection. Actually this is the case: it is
well-known (see e.g., \cite{husemoller}) that the set of equivariant
functions on a principal bundle taking values in some vector space $V$
is in bijective correspondence to the set of sections of the
associated vector bundle with fiber $V$. Since the structure group is discrete,
the bundle is always flat.

We therefore arrive at the conclusion that any unitary representation
of the mapping class group provides a vector bundle over $Q$ whose
space of square-integrable sections forms a possible Hilbert space for
geons in quantum gravity {\em as far as kinematics is
  concerned}. These spaces must have further imposed dynamical
constraints, i.e., they must be a suitable arena for dynamics before
they really can be claimed to be authentic quantum gravity Hilbert spaces
for geons. In this $(2 + 1)d$ context we can actually do more than
this. It is possible to impose all the dynamical constraints of
general relativity to obtain the reduced configuration space, thereby
taking into account dynamical aspects as well. Our space $Q$ will then
be the moduli space of the surface $\Sigma$, and its universal
covering $\tilde{Q}$ will be the Teichmuller space \cite{sum}. Again,
$\pi_1(Q)$ will be the mapping class group and our discussion will go mostly
along the same lines \footnote{Strictly speaking, this discussion is
valid for genus $g \geq 2$. For genus 1, things are a bit more
complicated. For details see, for instance, \cite{Louko}.}. This state of
affairs is somewhat reminiscent of
quantization of matter particles in Minkowski spacetime, in the
context of usual quantum mechanics. One can have many representations
of the Lorentz group, but on the one hand only some of them seem to be
realized in nature, and on the other hand the dynamics selects which
representation survives in each physical situation, e.g., the
tensorial representations for the Klein-Gordon field or the spinorial
and tensorial representations for the Dirac field.

Thus, we are led to the questions of whether, and how, the Hilbert
spaces we have presented in the previous Sections fit into this
scheme, and in particular whether the spin-statistics connection we
have found extends to these quantum gravity geon states. First of all,
we note that some of the Hilbert spaces ${\cal H}_r$, carrying
representations $r = ([a,b],\rho)$ of our field algebra
${\cal A}$ (in the notation defined in Section 5) carry
naturally a representation of the mapping class group: we recall that
the elements of this group were naturally included in ${\cal A}$,
 and hence we may say that $r$ contains a representation (possibly
 reducible) of the mapping class group. We will again denote this
 representation by $r$. Therefore the vector bundle
associated to $\tilde Q$ with fiber ${\cal H}_{r}$, where the
$r$'s are just as stated, gives another Hilbert space for geons
in quantum gravity, via its square-integrable sections. In this new
scenario the fibers are {\em internal} state spaces. 

However, we are now faced with another problem: do these sections carry a
representation of the mapping class group? In other words, is it
possible to extend the action on each fiber to an action on the space
of sections? The answer is known to be negative in general
\cite{Aneziris,extension}. One can
nevertheless implement operators corresponding to elements of
$M_{\Sigma}$ on states {\em localized} at a point $q \in Q$. In
particular, we may extend the operators ${\cal R}$ and $C_{2\pi}$, related to
statistics and spin respectively, to operators $\hat R$ and
$\hat{C}_{2\pi}$ acting on such localized
states. We will show at the end of this Section that such states obey
a spin-statistics connection inherited from the fibers. Here we
presently give a simple geometrical discussion to bring out the
reasoning behind these remarks. 

A very useful construction of the universal covering space $\tilde Q$
is as follows \cite{balachandran}. Let us assume that the space $Q$ is
connected (if this is not so we can always choose a connected
component). Let $q_0$ be a point of $Q$ which
once chosen is not to be 
changed. Let $\alpha_q$ be a path on $Q$ from $q_0$ to another point
$q \in Q$ \footnote{In this discussion all parametrized curves \{$\alpha:[0,1]
  \rightarrow Q | \alpha(0) = q_0; \alpha(1) = q$\} with different
  parametrizations but with same image in $Q$ are to be regarded as the
  same path.}. The path space ${\cal P}Q$ of $Q$ is the space
\{$\alpha_q$\} of these paths. Let us next say that two paths
$\alpha_q$ and $\alpha'_q$ are equivalent and write $\alpha_q \sim
\alpha'_q$ if one of them can be deformed to the other holding $q_0$
and $q$ fixed. One can then show that the space
$\tilde Q$ is the same as the space of equivalence classes
$[\alpha_q]$ of such paths, the projection map $p:{\tilde Q}
\rightarrow Q$ being given by $[\alpha_q] \mapsto q$. For a fixed $q$,
these equivalence classes form the fiber over the point $q$.

The group $\pi_1(Q)$ can be identified with the set of equivalence classes
of loops starting and ending at $q_0$, i.e., $[\alpha_{q_0}]$, with
the compositions of loops $[\alpha'_{q_0}][\alpha_{q_0}] =
[\alpha'_{q_0} \circ \alpha_{q_0}]$, where we first trace
$[\alpha_{q_0}]$, and then trace $[\alpha'_{q_0}]$. This group acts on
$\tilde Q$ on the right via the composition of paths:
$[\alpha_{q}][\alpha_{q_0}] = [\alpha_{q} \circ \alpha_{q_0}]$. This
corresponds to the free, fiberwise action of $\pi_1(Q)$ on $\tilde Q$
that we mentioned before. Note moreover that we have chosen a fiducial
point and put $\pi_1(Q) \simeq \pi_1(Q ; q_0)$, the latter denoting
the homotopy group of paths {\em based at $q_0$}, to be our ``model'' for
the structure group. Now, we might consider $\pi_1(Q ; q)$, the
homotopy group of paths based at another point $q \in Q$. This group is
isomorphic to $\pi_1(Q ; q_0)$, but the isomorphism is not canonical,
and we will see that this fact prevents the extension of the action of
$\pi_1(Q)$ when this group is non-abelian. We point out that $\pi_1(Q
; q)$ acts on the space \{$[\alpha_q]$\} on the left, and therefore
does not interfere with the action of $\pi_1(Q ; q_0)$; if we denote an
element of  $\pi_1(Q ; q)$ by $[\gamma_q]$, we have that
$[\gamma_q][\alpha_q] = [\gamma_q \circ \alpha_q]$. 

We now show that exchange and $2\pi$-rotation correspond to actions
like that of $\pi_1(Q; q)$, and not to the globally defined right
action of $\pi_1(Q)$. If $d$ denotes a diffeo and $h$ a metric, let
$hd$ denote the pull-back metric $d^{\ast}h$. Then $\tilde Q$
consists of elements $hDiff^{\infty}_0$, $Q$ of elements
$hDiff^{\infty}$, and $\pi_1(Q) = Diff^{\infty}/Diff^{\infty}_0$ acts
on $\tilde Q$ on the right \footnote{Let $G$ be any group and $X$ a space on
  which $G$ acts on the right. We may take the quotient $X/G$ of
  equivalence classes by the action. Let also $x \in X$. We denote by
  $xG$ the equivalence class of $x$ in $X/G$.}: $hDiff^{\infty}_0
\rightarrow (hDiff^{\infty}_0)(dDiff^{\infty}_0) = hdDiff^{\infty}_0$,
with $d \in Diff^{\infty}$. This action is globally defined. The
association of a $2\pi$-rotation or exchange however is not to this
action of $Diff^{\infty}/Diff^{\infty}_0$. Instead, it is obtained as
follows. Let $q = hDiff^{\infty} \in Q$ correspond to two
well-separated geons and let $\{q(t): 0 \leq t \leq 1\}$ be a loop in
$Q$ based at $q(0) = q(1) = q$. We can write $q(t) =
h(t)Diff^{\infty}$, and there exists an element $d \in Diff^{\infty}$ such
that $h(1) = h(0)d$, since $q(0) = q(1)$. Now it is shown elsewhere
\cite{Aneziris} that the physical process of exchange say is
associated with an element ${\cal R}Diff^{\infty}_0$ of
$Diff^{\infty}/Diff^{\infty}_0$ (${\cal R} \in Diff^{\infty}$) and a
loop based at $q(0) \in Q$. But this correspondence is not unique,
being deduced from the $\pi_1(Q ; q)$ action on $\tilde Q$. Thus according to
\cite{Aneziris}, the exchange process gives a curve $\{h(t)\}$ with
$h(1) = h(0){\cal R}$. It becomes the curve $\{ h(t)Diff^{\infty}_0
\}$ in $\tilde Q$ and the loop $\{ h(t)Diff^{\infty} \}$ in $Q$. Now
to find the diffeo and the loop for exchange based at $q(0)$, the
starting metric can be $h(0)$ or $\bar{h}(0) = h(0)d$, for any $d \in
Diff^{\infty}$ as both give rise to the same $q(0)$. Then for the two
curves $\{ h(t) \}$ and $\{ \bar{h}(t) \}$, we have $h(1) = h(0){\cal
  R}$ and $\bar{h}(1) = \bar{h}(0){\cal R}$. They give the curves $\{
h(t)Diff^{\infty}_0 \}$ and $\{ \bar{h}(t)Diff^{\infty}_0 \}$ in
$\tilde Q$ and the loops $\{ h(t)Diff^{\infty} \}$ and $\{
\bar{h}(t)Diff^{\infty} \}$ in $Q$. But {\em the homotopy classes of
  these loops are not in general the same, being related by the action
  of $\pi_1(Q ; q)$}. They do not therefore always define an
unambiguous element of $\tilde Q$. To see this first note that $\{
\bar{h}(t)d^{-1}Diff^{\infty}_0 \}$ also projects to the loop $\{
\bar{h}(t)Diff^{\infty} \}$  while at the same time it has the same
starting point $h(0)Diff^{\infty}_0$ as $\{ h(t)Diff^{\infty}_0
\}$. It follows that the lifts of the loops  $\{ h(t)Diff^{\infty} \}$ and $\{
\bar{h}(t)Diff^{\infty} \}$ to $\tilde Q$ with the same starting point
$h(0)Diff^{\infty}_0$ are $\{ h(t)Diff^{\infty}_0 \}$ and $\{
\bar{h}(t)d^{-1}Diff^{\infty}_0 \}$. But their endpoints in general are
different, being $h(0){\cal R}Diff^{\infty}_0$ and $h(0)d{\cal
  R}d^{-1}Diff^{\infty}_0$ respectively, showing that the two loops
may not be homotopic. Further the diffeos associated to the exchange
can be ${\cal R}$ or $d{\cal R}d^{-1}$ and their images in
$Diff^{\infty}/Diff^{\infty}_0$ can be different.

 Suppose then that we want to consider a global action of $\pi_1(Q)$
 on the quantum states. The two-geon configuration $hDiff^{\infty}
 \equiv q \in Q$ is described in quantum theory by an equivariant
 function $\Psi$ on $\tilde Q$ taking values in some vector space
 ${\cal H}_{\Gamma}$
 which carries a representation $\Gamma$ of
 $Diff^{\infty}/Diff^{\infty}_0$. We assume that $\Psi$ is localized
 at a point $[\alpha_q] \in \tilde{Q}$ in the fiber over $q$. The exchange
 process will then correspond to a loop class $[\gamma _q] \in \pi_1(Q
 ; q)$. This will act on the wave function by
\be 
\label{QG4}
([\gamma _q] \Psi)([\alpha _q]) = \Psi ([\gamma _q]^{-1}[\alpha _q]).
\ee
Since $[\gamma _q]^{-1}[\alpha _q] = [\gamma _{q}^{-1} \circ
\alpha _q]$ is in the fiber over $q$ as well, there exists a unique
$[\sigma ^{\gamma}_0] \in \pi_1(Q)$ such that $[\gamma _{q}^{-1} \circ
\alpha _q] = [\alpha _q][\sigma ^{\gamma}_0]$. Therefore
\be
\label{QG5}
 \Psi ([\gamma _q]^{-1}[\alpha _q]) = \Gamma ([\sigma
 ^{\gamma}_0]^{-1}) \Psi ([\alpha _q]).
\ee

However, this association of
an element of the fundamental group of $Q$ based at $q$ to an element
of $\pi_1(Q)$ 
is canonical only for $\Psi$ localized at a point $\tilde q$ in the
fiber over $q$ as stated, otherwise we
may pick another $[\alpha '_q] \in p^{-1}(q)$ which is related to
$[\alpha _q]$ by the relation $[\alpha '_q] = [\alpha _q][t]$, for some
$[t] \in \pi_1(Q)$. Then we have 
\bea
\label{QG6}
([\gamma _q] \Psi)([\alpha' _q]) = \Psi ([\gamma _q]^{-1}[\alpha' _q])
&=& \Psi ([\gamma _q]^{-1}[\alpha _q][t]) \\ \nonumber 
&=& \Psi ([\alpha _q][\sigma^{\gamma}_0][t]) 
= \Gamma ([t]^{-1}[\sigma ^{\gamma}_0]^{-1}[t]) \Psi
([\alpha' _q]).
\eea
Thus we have problems with continuity of the action for non-abelian
$\Gamma$ if we try to extend the action of $\pi_1(Q)$ to non-localized
states. 

We are now ready to consider the content of the spin-statistics
theorem in this context of localized states. Let ${\cal H}_{\Gamma}$ be a
Hilbert space carrying a representation $\Gamma$ of the algebra
for one geon. Consider the space of equivariant functions
$\Psi:\tilde{Q} \rightarrow {\cal H}_{\Gamma}$, which is the space of
states of the geon in quantum gravity, and denote it by
$V_{\Gamma}$. We have seen that the
mapping class group will not act on all of $V_{\Gamma}$, but will have
an action on localized states. Technically speaking these are ``delta
functions'' concentrated at some point $\tilde{q} \in \tilde{Q}$. This means
distributions, i.e., linear functionals $\delta_{\tilde{q}}$ on the
$\Psi$'s  such that, for each $\Psi \in
V_{\Gamma}$, we have
\be
\label{QG7}
 \delta _{\tilde q}(\Psi) \equiv \Psi(\tilde q).
\ee
Let $d \in Diff^{\infty}/Diff^{\infty}_0$, and $\tilde{q} =
hDiff^{\infty}_0$, where as before $h$ denotes some metric on
$\Sigma$. Then $Diff^{\infty}/Diff^{\infty}_0$ acts on the localized
space as follows:
\be
\label{QG8}
\hat{d} \delta_{\tilde q} = \delta_{\tilde {q} d^{-1}},
\ee
and given any state $\Psi \in V_{\Gamma}$, $ \delta_{\tilde {q} d^{-1}}(\Psi)
= \Psi(\tilde {q} d^{-1}) = \Gamma(d) \delta_{\tilde q}(\Psi)$, from
the equivariance property. Hence we may simply put, with a slight
abuse of notation, 
\be
\label{QG9}
\hat{d} \delta_{\tilde q} = \Gamma(d)\delta_{\tilde q}. 
\ee
We are actually interested in the action of ${\cal R}$
and $C_{2\pi}$. Like in the case of vectors in ${\cal H}_{\Gamma}$, we
can construct localized states with definite spin $s$, namely by
applying the spin projector $P_s$ in (\ref{spin1}) to any generic state.
Note, nevertheless that the definition given in
(\ref{QG8}) does not apply directly in this case, since $P_s$ is not an
element of the mapping class group as it has been defined. This needs
not bother us, though, since the definition of (\ref{QG9}) extends by
linearity to the algebra generated by the mapping class group, so we
may define a localized state of spin $s$,
denoted by $\delta_{\tilde q}^{(s)}$ through the equation
\be
\label{QG10}
\delta_{\tilde q}^{(s)} = \Gamma(P_s)\delta_{\tilde q}.
\ee
Now pick two equal states $\delta_{\tilde q}^{(s)}$ localized
at the same point $\tilde{q} \in \tilde{Q}$ and with the same spin $s$. These
are eigenvectors of the operator $\hat {C_{2\pi}}$ with eigenvalue
$\exp{i2\pi s}$, and we may write, for any two states $\Phi$ and
$\Psi$ in $V_{\Gamma}$:
\be
\label{QG11}
\hat{{\cal R}}~\delta_{\tilde q}^{(s)} \otimes \delta_{\tilde
  q}^{(s)}(\Phi,\Psi) =   
{\cal R}~\{\Gamma (P_s) \Phi(\tilde q) \otimes \Gamma(P_s) \Psi (\tilde
q)\}.
\ee
But since the vectors in the parenthesis are both
vectors in ${\cal H}_{\Gamma}$, the spin-statistics relation is valid
for them. Since $\Phi$ and $\Psi$ are arbitrary, we may conclude from
(\ref{ssc}) that
\be 
\label{QG12}
\hat{{\cal R}}~\delta_{\tilde q}^{(s)} \otimes \delta_{\tilde q}^{(s)} = 
e^{i(2\pi s - \theta _r)}~\delta_{\tilde q}^{(s)} \otimes
\delta_{\tilde q}^{(s)},
\ee
which is the expression for the spin-statistics connection for the
localized geon states in quantum gravity.

\sxn{Final Remarks}\label{S8}

In this paper, we have shown how to explicitly encode information on
non-trivial spatial topology in the quantum theory of topological
geons. We have described its
classical degrees of freedom by using the low-energy limit of a
Yang-Mills theory coupled to a Higgs field in the Higgs phase, where
the symmetry is spontaneously broken down to a finite gauge group. It
has been argued that this is enough to capture aspects of the topology of the
underlying space manifold, and it has been shown how such a theory can
be quantized. The field algebra for one or many geons has
been derived, and its representations, corresponding to the various geonic
sectors of the theory, have been worked out in detail. 

Our discussion borrowed heavily from the theory of  quantization of vortices
\cite{prop}. It led to a striking consequence: we have been
able to derive a new spin-statistics relation obeyed by the geonic
states. In this relation, there is still a parameter $\theta _r$ to be
fixed for each
representation of the geon algebra ${\cal A}^{(1)}$, but to obtain it
explicitly one has to to fix a finite group $H$ and use the formalism
to work out the representations of the geon algebra. We will attempt
this elsewhere. We have also
shown how the geonic states we describe here correspond to quantum
gravity states of geons.

One is naturally led to inquire whether the framework developed in
this paper can be extended to cover the more general case when the
gauge group is a
Lie group. In particular it is known that general relativity in $(2 +
1)d$ can be viewed as a Chern-Simons theory with gauge group
$ISO(2,1)$ \cite{iso}, and therefore one may investigate the spin-statistics
connection when $H$ in this paper is replaced by $ISO(2,1)$. Such
generalizations will be the issues of a forthcoming paper.

\vspace{.5cm}
\noindent
{\large \bf Acknowledgments}

\noindent
We would like to thank J.C.A. Barata, B.G.C. da Cunha, A. Momen and S. Vaidya
for many helpful discussions. The work of A.P. Balachandran was
supported by the Department of
Energy, U.S.A., under contract number DE-FG02-85ERR40231. The work of
E. Batista, I.P. Costa e Silva and
P. Teotonio-Sobrinho was supported by FAPESP, CAPES and CNPq
respectively. A.P.B. acknowledges the wonderful hospitality of Denjoe
O'Connor and support of CINVESTAV at Mexico, D.F. while this work was
being completed.

\end{document}